\documentclass[aip,11pt,reprint,floatfix,superscriptaddress,longbibliography]{revtex4-2}
\usepackage[T1]{fontenc}
\usepackage[utf8]{inputenc}
\usepackage{multirow}
\usepackage{amsmath}
\usepackage{graphicx}
\usepackage{bm}
\usepackage{braket}
\usepackage{soul} 
\usepackage{tcolorbox}
\usepackage{url}
\usepackage[linktocpage, colorlinks=true, linkcolor=darkblue, citecolor=darkblue, urlcolor=darkblue]{hyperref}
\definecolor{darkblue}{rgb}{0,0,.5}

\usepackage[normalem]{ulem}

\begin{document}

\preprint{AIP/123-QED}

\title[]{Quantum-Electrodynamical Time-Dependent Density Functional Theory. I. A Gaussian Atomic Basis Implementation}

\author{Junjie Yang}
\affiliation{Department of Chemistry and Biochemistry, University of Oklahoma, Norman, Oklahoma 73019, USA}

\author{Qi Ou}
\email{qiou@tsinghua.edu.cn}
\affiliation{MOE Key Laboratory of Organic OptoElectronics and Molecular Engineering, Department of Chemistry, Tsinghua
University, Beijing 100084, China.}

\author{Zheng Pei}
\affiliation{State Key Laboratory of Physical Chemistry of Solid Surfaces, Collaborative Innovation Center of Chemistry for Energy Materials, 
Fujian Provincial Key Laboratory of Theoretical and Computational Chemistry, and Department of Chemistry, College of Chemistry and Chemical Engineering, Xiamen University, Xiamen 361005, P. R. China.}

\author{Hua Wang} 
\affiliation{Homer L. Dodge Department of Physics and Astronomy, University of Oklahoma, Norman, Oklahoma 73019, USA} 

\author{Binbin Weng} 
\affiliation{Microfabrication Research and Education Center and School of Electrical and Computer Engineering, University of Oklahoma, Norman, Oklahoma 73019, USA}

\author{Zhigang Shuai}
\email{zgshuai@tsinghua.edu.cn}
\affiliation{MOE Key Laboratory of Organic OptoElectronics and Molecular Engineering, Department of Chemistry, Tsinghua
University, Beijing 100084, China.}

\author{Kieran Mullen}
\email{kieran@ou.edu}
\affiliation{Homer L. Dodge Department of Physics and Astronomy, University of Oklahoma, Norman, Oklahoma 73019, USA} 

\author{Yihan Shao}
\email{yihan.shao@ou.edu}
\affiliation{Department of Chemistry and Biochemistry, University of Oklahoma, Norman, Oklahoma 73019, USA}

\date{\today}

\begin{abstract}
Inspired by the formulation of quantum-electrodynamical time-dependent density functional theory (QED-TDDFT) by Rubio and coworkers, we propose an implementation that uses dimensionless amplitudes for describing the photonic contributions to QED-TDDFT electron-photon eigenstates. The leads to a symmetric QED-TDDFT coupling matrix, 
which is expected to facilitate the future development of analytic derivatives.  Through a Gaussian atomic basis implementation of the QED-TDDFT method, we examined the effect of dipole self-energy, rotating wave approximation, and the Tamm-Dancoff approximation on the QED-TDDFT eigenstates of model compounds (ethene, formaldehyde, and benzaldehyde) in an optical cavity.  We highlight, in the strong coupling regime, the role of higher-energy and off-resonance excited states with large transition dipole moments in the direction of the photonic field, which are automatically accounted for in our QED-TDDFT calculations and might substantially affect the energy and composition of polaritons associated with lower-energy electronic states. 
\end{abstract}

\maketitle

\section{\label{sec:Introduction}Introduction}

Quantum optics effects on atoms have been extensively studied during the last several decades,\cite{meschede_radiating_1992,raimond_manipulating_2001,mabuchi_cavity_2002,walther_cavity_2006,gleyzes_quantum_2007,englund_controlling_2007,guerlin_progressive_2007,ruggenthaler_quantum-electrodynamical_2018,head-marsden_quantum_2021} enabling scientists to shift atomic energy levels,\cite{lutken_energy-level_1985,barton_quantum-electrodynamic_1987} tune atomic electronic transition rates,\cite{gabrielse_observation_1985,hulet_inhibited_1985,jhe_suppression_1987} and generate quantum systems with atom-atom entangled states.\cite{hagley_generation_1997,maitre_quantum_1997} In contrast, the behavior of molecules in optical cavities attracted a lot of attention only in recent years.\cite{drexhage_iv_1974,fujita_tunable_1998,lidzey_strong_1998,dintinger_strong_2005,ebbesen_hybrid_2016,frisk_kockum_ultrastrong_2019,herrera_molecular_2020}
In particular, several molecules have been shown to couple strongly to a quantized radiation field, causing their electronic states to hybridize with the cavity photon levels to produce superpositions and entanglements.\cite{raimond_manipulating_2001,triana_entangled_2018,gu_manipulating_2020}
The study of such entangled states lead to the establishment of the field of polariton chemistry, which focuses on the use of optical cavities to manipulate chemical and photochemical reactivities,\cite{schwartz_reversible_2011,hutchison_modifying_2012,fontcuberta_i_morral_ultrastrong_2012,hutchison_tuning_2013,canaguier-durand_thermodynamics_2013,salomon_strong_2013,wang_phase_2014,shalabney_coherent_2015,george_liquid-phase_2015,thomas_ground-state_2016,kowalewski_manipulating_2017,stranius_selective_2018,galego_cavity_2019,eizner_inverting_2019,eizner_inverting_2019,herrera_molecular_2020} modify the intersystem crossing rates,\cite{hutchison_modifying_2012,fontcuberta_i_morral_ultrastrong_2012,stranius_selective_2018,ulusoy_modifying_2019,takahashi_singlet_2019,eizner_inverting_2019} and enhance organic molecule light emitting efficiencies.\cite{kena-cohen_room-temperature_2010,schwartz_polariton_2013,wang_quantum_2014,george_ultra-strong_2015,shalabney_enhanced_2015,orgiu_conductivity_2015,barachati_tunable_2018,held_ultrastrong_2018,jayaprakash_hybrid_2019}


In principle, a coupled molecule-photon system is best described by the relativistic quantum field theory (QFT).\cite{ruggenthaler_time-dependent_2011}
But to avoid the computational complexity of QFT, many non-relativistic and simplified theories have been developed.\cite{mandal_polariton-mediated_2020} Within the Rabi model, for instance, one adopts a semiclassical approach that combines a non-relativistic quantum mechanical description for the molecule and a classical description of the electromagnetic field.\cite{rabi_process_1936,rabi_space_1937,braak_integrability_2011,xie_quantum_2017} The Pauli-Fierz model was introduced to provide a consistent treatment of the spontaneous emission.\cite{pauli_zur_1938,bethe_electromagnetic_1947} Jaynes and Cummings proposed two other similar quantum models, which are known as the Jaynes-Cummings (JC) model and rotating-wave approximation (RWA).\cite{jaynes_comparison_1963,shore_jaynes-cummings_1993}  Within both models, the counter-rotating terms (CRT) are neglected, which turns out to be a valid approximation for the resonance and near-resonance conditions and weak coupling regimes.\cite{jaynes_comparison_1963,shore_jaynes-cummings_1993,mandal_polariton-mediated_2020}

 
In the study of polariton chemistry, it has been common to restrict the description of each electronic system to simplified two- or three{-}state model.\cite{braak_integrability_2011,garraway_dicke_2011,xie_quantum_2017,de_bernardis_breakdown_2018,schafer_ab_2018,ribeiro_polariton_2018,chen_ehrenfestr_2019-1,chen_ehrenfestr_2019,li_quasiclassical_2020,herrera_molecular_2020,mordovina_polaritonic_2020} The corresponding input parameters (energy levels and coupling elements), on which the models would heavily rely, are obtained from either experiments or first-principles quantum chemistry calculations. Through employing these models, it has been predicted that the coupling of molecular systems to quantized radiation modes could substantially modify the potential energy surfaces and create new conical intersections,\cite{hutchison_modifying_2012,fontcuberta_i_morral_ultrastrong_2012,galego_many-molecule_2017,ulusoy_modifying_2019} 
suppress or enhance photoisomerization reactions,\cite{litinskaya_fast_2004,michetti_simulation_2008,canaguier-durand_non-markovian_2015,orgiu_conductivity_2015,galego_suppressing_2016,herrera_absorption_2017} increase the charge transfer and excitation energy transfer rates,\cite{feist_extraordinary_2015,zhong_non-radiative_2016,herrera_cavity-controlled_2016,du_theory_2018,semenov_electron_2019,mandal_polariton-mediated_2020} accelerate the singlet fission kinetics,\cite{martinez-martinez_polariton-assisted_2018,takahashi_singlet_2019} and even control the chemical reactions remotely.\cite{thanopulos_laser-driven_2004,stranius_selective_2018} These observations open up opportunities to use optical cavities to manipulate chemical and photochemical reactions. 

    
Recently, \emph{ab initio} quantum mechanical frameworks were proposed to describe interacting electrons and photons.\cite{ruggenthaler_time-dependent_2011,tokatly_time-dependent_2013,ruggenthaler_quantum-electrodynamical_2014,pellegrini_optimized_2015,trevisanutto_hedin_2015,flick_kohnsham_2015,kowalewski_non-adiabatic_2016,de_melo_unified_2016,flick_atoms_2017,schafer_ab_2018,fregoni_manipulating_2018,flick_lightmatter_2019,fregoni_strong_2020,flick_ab_2020} Specifically, Rubio and others developed quantum electrodynamical density functional theory (QEDFT)\cite{tokatly_time-dependent_2013,ruggenthaler_quantum-electrodynamical_2014,pellegrini_optimized_2015,flick_kohnsham_2015,flick_cavity_2017,flick_atoms_2017,tokatly_conserving_2018,flick_ab_2018,flick_ab_2020} and quantum electrodynamics coupled-cluster (QED-CC) theory,\cite{haugland_coupled_2020,deprince_cavity-modulated_2021,haugland_intermolecular_2021} which inherit an accurate description of molecular electronic structure from time-dependent density functional theory (TDDFT), coupled-cluster theory (CC), equation-of-motion coupled-cluster theory (EOM-CC), and other \textit{ab initio} electronic structure theories.

In this work, we closely follow the QEDFT method from Rubio and coworkers.\cite{tokatly_time-dependent_2013,ruggenthaler_quantum-electrodynamical_2014,pellegrini_optimized_2015,flick_kohnsham_2015,flick_cavity_2017,flick_atoms_2017,tokatly_conserving_2018,flick_ab_2018,flick_lightmatter_2019,flick_ab_2020} 
Through a slightly modified matrix formulation of linear-response quantum electrodynamical time-dependent density functional theory (QED-TDDFT), we obtain a symmetric TDDFT-Pauli-Fierz (TDDFT-PF) Hamiltonian for coupling molecules and cavity photons.  This allows us to systematically examine the effects of CRT, dipole self-energy (DSE), and the Tamm-Dancoff approximation (TDA), and to compare the polariton energies to different model Hamiltonian results.  We expect that, as the electron-photon coupling strength increases, there might be (a) a substantial deviation from symmetric Rabi splitting (which arises from the 2-state model Hamiltonian) and (b) significant differences in the polariton energy and compositions among various theoretical models. 

This article is organized as follows. The TDDFT-PF formula is derived in Section~\ref{sec::qedtddft} and in Appendix \ref{appendix::lr} and \ref{appendix::eom} (with linear-response and equation-of-motion formulations, respectively), with its no-DSE, RWA, and TDA variations presented in Section~\ref{sec::prism}. 
Section~\ref{sec::detail} describes our Gaussian atomic basis implementation within the \textsc{PySCF} software package.\cite{sun_pyscf_2018}
Preliminary results on the polariton states of ethene, formaldehyde, and benzaldehyde molecules in the optical cavity are reported in Section~\ref{sec::results}, which is followed by concluding remarks in Section~\ref{sec::conclusions}. 


\section{\label{sec:Theory}Theory}

\subsection{Notation}
For a molecule, we will use indices $i,j$ to represent its occupied Kohn-Sham orbitals, and $a,b$ to denote its unoccupied (virtual) Kohn-Sham orbitals.  The corresponding orbital energies will be written as $\varepsilon_i, \varepsilon_j, \varepsilon_a$, and $\varepsilon_b$.  The dipole moment vector in the basis of these orbitals is 
\begin{equation}
\bm{\mu}_{ai} = \bigg( \left< a \middle| \hat{x} \middle| i \right>,
\left< a \middle| \hat{y} \middle| i \right>, 
\left< a \middle| \hat{z} \middle| i \right> \bigg)
\end{equation}
$\mathbf{A}$ and $\mathbf{B}$ refer to conventional TDDFT coupling super-matrices, while $\mathbf{X}$ and $\mathbf{Y}$ are TDDFT amplitudes.\cite{casida_time-dependent_1995,bauernschmitt_treatment_1996,hirata_time-dependent_1999,fadda_time-dependent_2003,dreuw_single-reference_2005,casida_time-dependent_2009,casida_progress_2012,chen_random-phase_2017}

For an optical cavity with $M$ photon modes, its $\alpha$-th mode of frequency is denoted by $\omega_\alpha$. The corresponding fundamental coupling strength \cite{schafer_ab_2018,flick_lightmatter_2019}
\begin{equation}
\bm{\lambda}_\alpha = \sqrt{\frac{1}{\epsilon_0}} S_\alpha(\boldsymbol{r}_0) \boldsymbol{\epsilon}_\alpha,\quad \quad \alpha=1, 2, \cdots, M     
\label{eq:lambda_alpha} 
\end{equation}
depends on the transversal polarization vector $\boldsymbol{\epsilon}_\alpha$.
For a Fabry-P\'{e}rot cavity of volume $V_{\textrm{tot}}=L_xL_yL_z$ and with mirror planes perpendicular to the $z$-axis, the dimensionless mode function  
\begin{equation}
S_\alpha(\boldsymbol{r}) =  \sqrt{\frac{2}{V_{\textrm{tot}}}} \textrm{sin} \left( \frac{\alpha \pi z} {L_z} \right)       \label{eq:S_alpha} 
\end{equation}
is evaluated at a chosen reference point $\boldsymbol{r}_0$ for the molecular subsystem.  When there are $N$ identical and non-interacting molecules with the same orientation, the effective volume of each molecule, $V_\mathrm{eff}=\frac{V_{\textrm{tot}}}{N}$, can be used in place of $V_{\textrm{tot}}$ in Eq. \ref{eq:S_alpha}.\cite{hutchison_modifying_2012}  

For each photon mode, the corresponding displacement coordinate and conjugate moment refer to 
\begin{eqnarray}
\hat{q}_\alpha & =  & \sqrt{ \frac{\hbar}{2 \omega_\alpha}} \left( \hat{b}_\alpha + \hat{b}_\alpha^\dagger \right) \label{eq:q_alpha}  \\
\hat{p}_\alpha & = &  - i \sqrt{ \frac {\hbar \omega_\alpha}{2}} \left( \hat{b}_\alpha - \hat{b}_\alpha^\dagger \right) \label{eq:p_alpha} 
\end{eqnarray}
with $\hat{b}_\alpha^\dagger$ and $\hat{b}_\alpha$ being the creation and annihilation operators for the mode.  For convenience, the dot product of the dipole moment of each virtual-occupied pair and the photon field will be written as 
\begin{eqnarray}
\lambda_{ai}^\alpha = \bm{\mu}_{ai} \cdot \bm{\lambda}_\alpha 
\end{eqnarray} 

\subsection{\label{sec::qedtddft}QED-TDDFT Equation within the Pauli-Fierz Hamiltonian}

For a molecule confined in this optical cavity, by making the Born-Oppenheimer approximation and the long-wavelength or dipole approximation in the length-gauge, its Pauli-Fierz Hamiltonian can be written as\cite{flick_lightmatter_2019,mandal_polariton-mediated_2020}
\begin{eqnarray}
\hat{H} & =  &  \hat{H}_\mathrm{elec} + \sum_{\alpha=1}^M \left[ \frac{1}{2} \hat{p}_{\alpha}^2 
+  \frac{1}{2} \omega_\alpha^2 \left( \hat{q}_\alpha - \frac{ 1} { \omega_\alpha}  \bm{\lambda}_\alpha \cdot \left< \hat{\bm{\mu}} \right> \right)^2 \right] \nonumber \\
& & 
+ \sum_{\alpha=1}^M \frac{ j_\alpha(t) } { \omega_\alpha} \hat{q}_\alpha \label{eq:Pauli-Fierz} 
\end{eqnarray}
where the photon modes interact with the molecular dipole moment, $\left< \hat{\bm{\mu}} \right>$, and $j_\alpha(t)$ is the source of the $\alpha$-th cavity mode.

If the electronic Hamiltonian, $\hat{H}_\mathrm{elec}$, is described by the Kohn-Sham density functional theory, the corresponding TDDFT-PF equation is shown in the Appendix \ref{appendix::lr} and Appendix \ref{appendix::eom} to be,
\begin{eqnarray}
& \begin{bmatrix}
\mathbf{A}+\bm{\Delta} & \mathbf{B}+\bm{\Delta} & \hbar \mathbf{g}^\dagger & \hbar \tilde{\mathbf{g}}^\dagger \\
\mathbf{B}+\bm{\Delta} & \mathbf{A}+\bm{\Delta} & \hbar \mathbf{g}^\dagger &  \hbar \tilde{\mathbf{g}}^\dagger  \\
\hbar \mathbf{g} & \hbar \mathbf{g} & \hbar \boldsymbol{\omega} & 0 \\
\hbar \tilde{\mathbf{g}} & \hbar \tilde{\mathbf{g}}  & 0 & \hbar \boldsymbol{\omega}  
\end{bmatrix}
\begin{bmatrix} 
\mathbf{X} \\ \mathbf{Y} \\ \mathbf{M} \\  \mathbf{N}
\end{bmatrix} \nonumber \\
& \quad \quad \quad \quad = \hbar \Omega^{\textrm{TDDFT-PF}} 
\begin{bmatrix}
\mathbf{1} &0 & 0 & 0  \\
0 & -\mathbf{1}  & 0 & 0  \\
0 & 0 & \mathbf{1}  & 0 \\
0 & 0 & 0 & -\mathbf{1}   
\end{bmatrix}
\begin{bmatrix} 
\mathbf{X} \\ \mathbf{Y} \\ \mathbf{M} \\  \mathbf{N}
\end{bmatrix}
\label{eq:tddft-pf}
\end{eqnarray}

Here, the electron-electron block contains TDDFT supermatices, $\mathbf{A}$ and $\mathbf{B}$, as augmented by the DSE terms,\cite{mandal_polariton-mediated_2020,haugland_intermolecular_2021}
\begin{equation}
\Delta_{ai,bj}  =   \sum_{\alpha=1}^M   \lambda_{ai}^\alpha \lambda_{bj}^\alpha
\label{eq:delta-aibj} 
\end{equation}
while the electron-photon and photon-electron blocks are
\begin{equation}
   \hbar g_{bj}^\alpha = \hbar \tilde{g}_{bj}^\alpha =   \sqrt{\frac{\hbar \omega_\alpha}{2}}  \lambda_{bj}^\alpha
\end{equation}
whereas the photon-photon block is a diagonal matrix,
\begin{equation}
    \omega_{\alpha\beta} = \delta_{\alpha\beta} \omega_\alpha
\end{equation}


Due to the couplings (\emph{i.e.}, the electron-photon and photon-electron blocks), the $I$-th eigenvector of the TDDFT-PF equation, $(\mathbf{X}^I, \mathbf{Y}^I, \mathbf{M}^I, \mathbf{N}^I)^\mathrm{T}$, contains photonic amplitudes ($\mathbf{M}^I$ and $\mathbf{N}^I$) in addition to normal electronic amplitudes ($\mathbf{X}^I$ and $\mathbf{Y}^I$).
The ortho--normalization condition is,
\begin{equation}
    \sum_{ai} \left( X_{ai}^I X_{ai}^J - Y_{ai}^I Y_{ai}^J \right) 
    + \sum_{\alpha} \left( M_\alpha^I M_\alpha^J - N_\alpha^I N_\alpha^J \right) = \delta_{IJ}
\end{equation}
These components could be explained to be the Fourier coefficients of the time-dependent parameters\cite{scuseria_particle-particle_2013,ziegler_derivation_2014,yang_analysis_2020} (also see Appendices \ref{appendix::lr} and \ref{appendix::eom}).

\subsection{\label{sec::prism}A Prism of QED-TDDFT Methods}

\begin{figure}[htp]
    \centering
    \includegraphics[width=0.45\textwidth]{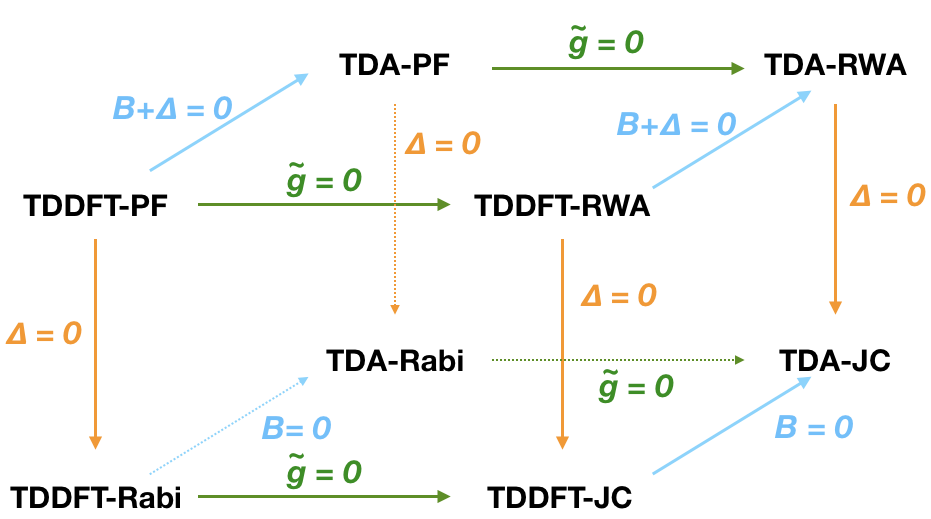}
    \caption{A prism of QED-TDDFT methods.}
    \label{fig:prism_methods}
\end{figure}


The TDDFT-PF formula (Eq. \ref{eq:tddft-pf}) can be approximated in a number of ways shown in Fig. \ref{fig:prism_methods}.  First, one can invoke the Tamm-Dancoff approximation (TDA),\cite{hirata_time-dependent_1999} which sets the $\mathbf{B}+\bm{\Delta}$ elements to zero.  One then  obtains the TDA-PF model,
\begin{align}
\begin{bmatrix}
\mathbf{A}+\bm{\Delta}  & \hbar \mathbf{g}^\dagger & \hbar \tilde{\mathbf{g}}^\dagger \\
\hbar \mathbf{g}  & \hbar \boldsymbol{\omega} & 0 \\
\hbar \tilde{\mathbf{g}}  & 0 & \hbar \boldsymbol{\omega}  
\end{bmatrix}
\begin{bmatrix} 
\mathbf{X} \\ \mathbf{M} \\  \mathbf{N}
\end{bmatrix} 
= \hbar \Omega^{\textrm{TDA-PF}} 
\begin{bmatrix}
\mathbf{1} & 0 & 0  \\
0 & \mathbf{1}  & 0 \\
0 & 0 & -\mathbf{1}   
\end{bmatrix}
\begin{bmatrix} 
\mathbf{X} \\ \mathbf{M} \\  \mathbf{N}
\end{bmatrix}
\label{eq:tda-pf} 
\end{align} 

If the DSE addition, $\bm{\Delta}$, to matrix $\mathbf{A}$ is further removed, one obtains the TDA-Rabi model,\cite{mandal_investigating_2019}
\begin{align}
\begin{bmatrix}
\mathbf{A}  & \hbar \mathbf{g}^\dagger & \hbar \tilde{\mathbf{g}}^\dagger \\
\hbar \mathbf{g}  & \hbar \boldsymbol{\omega} & 0 \\
\hbar \tilde{\mathbf{g}}  & 0 & \hbar \boldsymbol{\omega}  
\end{bmatrix}
\begin{bmatrix} 
\mathbf{X} \\ \mathbf{M} \\  \mathbf{N}
\end{bmatrix} 
= \hbar \Omega^{\textrm{TDA-Rabi}} 
\begin{bmatrix}
\mathbf{1} & 0 & 0  \\
0 & \mathbf{1}  & 0 \\
0 & 0 & -\mathbf{1}   
\end{bmatrix}
\begin{bmatrix} 
\mathbf{X} \\ \mathbf{M} \\  \mathbf{N}
\end{bmatrix}
\label{eq:tda-rabi} 
\end{align} 

If CRT terms (equal to $\hbar \tilde{\mathbf{g}}^\dagger$) are neglected from Eq. \ref{eq:tda-pf}, it amounts to the RWA approximation,\cite{mandal_investigating_2019}
\begin{align}
\begin{bmatrix}
\mathbf{A}+\bm{\Delta}  & \hbar \mathbf{g}^\dagger  \\
\hbar \mathbf{g}  &\hbar \boldsymbol{\omega}
\end{bmatrix}
\begin{bmatrix} 
\mathbf{X} \\ \mathbf{M} 
\end{bmatrix} 
= \hbar \Omega^{\textrm{TDA-RWA}} 
\begin{bmatrix}
\mathbf{1} & 0  \\
0 & \mathbf{1} 
\end{bmatrix}
\begin{bmatrix} 
\mathbf{X} \\ \mathbf{M} 
\end{bmatrix}
\label{eq:tda-rwa} 
\end{align} 

Finally, if both DSE and CRT terms are neglected in the TDA-PF formula (Eq. \ref{eq:tda-pf}), one arrives at the TDA-Jaynes-Cummings (TDA-JC) model
\begin{align}
\begin{bmatrix}
\mathbf{A}  & \hbar \mathbf{g}^\dagger  \\
\hbar \mathbf{g}  & \hbar \boldsymbol{\omega}
\end{bmatrix}
\begin{bmatrix} 
\mathbf{X} \\ \mathbf{M} 
\end{bmatrix} 
= \hbar \Omega^{\textrm{TDA-JC}} 
\begin{bmatrix}
\mathbf{1} & 0  \\
0 & \mathbf{1} 
\end{bmatrix}
\begin{bmatrix} 
\mathbf{X} \\ \mathbf{M} 
\end{bmatrix}
\label{eq:tda-jc} 
\end{align} 

Clearly, the first-order DSE correction to the energy of the $I$-th TDA-JC polariton is always positive
\begin{eqnarray}
\Omega_I^{\textrm{TDA-RWA}}  = \Omega_I^{\textrm{TDA-JC}}
+\sum_{\alpha=1}^M   \left| \bm{\lambda}_\alpha \cdot \bm{\mu}_I \right|^2 + \cdots
\label{eq::dse}
\end{eqnarray}
with the transition dipole moment of the $I$-th polariton being
\begin{eqnarray}
\bm{\mu}_I^{\textrm{TDA-JC}} = \sum_{ai} X_{I,ai}^{\textrm{TDA-JC}} \bm{\mu}_{ai} 
\end{eqnarray}

In contrast, the leading CRT correction to the TDA-JC energy is second-order and always negative 
\begin{align}
\Omega_I^{\textrm{TDA-Rabi}}  = \Omega_I^{\textrm{TDA-JC}}
- \sum_{\alpha=1}^M \frac{ g_{\alpha,I}^2}{ \Omega_I^{\textrm{TDA-JC}} + \omega_{\alpha}}  + \cdots
\label{eq::crt}
\end{align}
where 
\begin{eqnarray}
 g_{\alpha,I} = \sqrt{\frac{ \omega_\alpha}{2 \hbar}}  \left( \bm{\lambda}_\alpha \cdot \bm{\mu}_I^{\textrm{TDA-JC}} \right)  
 \label{eq:g_alpha_I} 
\end{eqnarray}
is the coupling between the $I$-th excited state and the $\alpha$-th cavity mode.
Note that under the resonance condition ($\Omega_I^{\textrm{TDA-JC}} = \omega_{\alpha}$) as well as $g_{\alpha,I}/\omega_{\alpha} \ll 1$, the leading CRT contributions are second-order to $g_{\alpha,I}$ (and thus $\lambda_\alpha$).  In those cases, the CRT contribution is roughly $\frac{1}{4}$ of the DSE term, but with an opposite sign.  This is different from the report from Huo and coworkers \cite{stokes_gauge_2019,mandal_polariton-mediated_2020} who found the CRT contribution to be $-\frac{1}{2}$ times the DSE term.  This discrepancy will be explained later in the context of QED-TDDFT results on test molecules. 

As shown in Fig. \ref{fig:prism_methods}, the TDDFT-Rabi, TDDFT-RWA, and TDDFT-JC models can be defined in a similar procedure starting from the TDDFT-PF model.  For instance, the TDDFT-JC working equation would be
\begin{align}
\begin{bmatrix}
\mathbf{A}  & \mathbf{B} & \hbar \mathbf{g}^\dagger \\
\mathbf{B} & \mathbf{A} & \hbar \mathbf{g}^\dagger \\
\hbar \mathbf{g}  & \hbar \mathbf{g} & \hbar \boldsymbol{\omega}  
\end{bmatrix}
\begin{bmatrix} 
\mathbf{X} \\ \mathbf{Y} \\  \mathbf{M}
\end{bmatrix} 
= \hbar \Omega^{\textrm{TDDFT-JC}} 
\begin{bmatrix}
\mathbf{1} & 0 & 0  \\
0 & \mathbf{-1}  & 0 \\
0 & 0 & \mathbf{1}   
\end{bmatrix}
\begin{bmatrix} 
\mathbf{X} \\ \mathbf{Y} \\  \mathbf{M}
\end{bmatrix}
\label{eq:tddft-jc} 
\end{align} 
Clearly, if this equation (or its TDA-JC counterpart in Eq. \ref{eq:tda-jc}) is recast in the representation of gas-phase TDDFT (or TDA) eigenstates, 
it is naturally reduced to the familiar Jaynes-Cummings model, but coupling all excited states to the photon field.

\section{\label{sec::detail}Computational Details}
The QED-TDDFT methods are implemented in a modified version of the \textsc{PySCF} software package\cite{sun_pyscf_2018} 
as well as the  {\sc Q-Chem} software package.\cite{shao_advances_2015} 
Several functionals (such as PBE,\cite{perdew_generalized_1996} PBE0,\cite{adamo_toward_1999}
B3LYP,\cite{becke_density-functional_1988,becke_new_1993,lee_development_1988}
and $\omega$B97X-D\cite{chai_long-range_2008}) are supported in our implementation. 
Only PBE0 results are presented in the next section,  
while the use of other functionals is found to lead to qualitatively similar results for the test systems. 
The lowest eigenstates are solved using Davidson's diagonalization algorithm.\cite{davidson_iterative_1975}

The ground-state geometries of all the molecules are obtained at the PBE0/6-311++G** level of theory using the {\sc Q-Chem} software package.
Planar Fabry-P\'{e}rot micro-cavities are chosen with a frequency resonant (or near-resonant) with the first excited states with a significant oscillation strength of each molecule.
Namely, a single fundamental coupling strength vector ($\bm{\lambda}$) is set to be parallel to the transition dipole moment of that particular excited state. 
The coupling strength is tuned by varying the concentration, while the maximum coupling strength is obtained using the estimated volume of each molecule. Note that this assumes all molecules have exactly the same orientation in the optical cavity. 
The coupling strengths $\lambda$ are represented in atomic unit, as $1 \, \mathrm{au} = 1 \sqrt{m}_\mathrm{e} E_\mathrm{h}/e\hbar$.  

Furthermore, we consider only one mode of the radiation field with $\alpha$=1 and apply the long wavelength approximation by setting $z$ to half way between the two mirrors.  
In the end, $S_\alpha(\boldsymbol{r})$ in Eq. \ref{eq:S_alpha} has a value of $\sqrt{\frac{2}{V_{\textrm{eff}}}}$ in all our calculations.

Three molecular systems are considered in this work:
\begin{itemize}
    \item For the ethene molecule, the effective molecular volume is estimated to be $2331 \, a_0^3$, which corresponds to a maximum coupling strength of $\lambda_\mathrm{max} = 0.1038 \, \mathrm{au}$. The cavity frequency is set to be $6.961$ eV, which is resonant with the first TDA excited state in vacuum; and the coupling strength vector ($\bm{\lambda}$) is parallel to the corresponding electronic transition dipole moment. 
    \item For the formaldehyde molecule, the effective molecular volume is estimated to be $1991 \, a_0^3$, which amounts to a maximum coupling strength of $\lambda_\mathrm{max} = 0.1123$ au.
    The cavity frequency is set to be $ 6.777$ eV in the TDDFT calculations and $6.784$ eV in the TDA calculations, which is in resonant with the second excited state from corresponding gas--phase calculations. 
    \item For the benzaldehyde molecule, the effective molecular volume is estimated to be $73050 \, a_0^3$, with the corresponding maximum coupling strength being $\lambda_\mathrm{max} = 0.0185 \, \mathrm{au}$. The cavity frequency is set to be either in resonance ($4.879 \, \mathrm{eV}$ for TDA while $4.810 \, \mathrm{eV}$ for TDDFT) with or $0.02 \, \mathrm{eV}$ off-resonant ($4.899 \, \mathrm{eV}$ for TDA and $4.830 \, \mathrm{eV}$ for TDDFT) from the second excitation energy.
\end{itemize}

\section{Results and Discussions}
\label{sec::results}

In this section, some preliminary results are presented. In Subsection~\ref{sec::c2h4}, the ethene molecule (C$_2$H$_4$) is used as a model system in resonance with the cavity mode to show the equivalence of Jaynes-Cummings model and the corresponding TDA-JC method. In Subsection~\ref{sec::ch2o}, polariton spectra of the formaldehyde molecule (also in resonance with the photon field) are shown at various levels of QED-TDDFT models to systematically analyze the effects of DSE and CRT terms as well as the TDA approximation.
In Subsection~\ref{sec::c6h5cho}, the benzaldehyde molecule is  studied as a more practical example of photochrome. The TDA-JC results are compared to those based on the two--state model, where the cavity frequencies are set to be in resonance or off-resonance with the first TDA excitation energy.

\subsection{\label{sec::c2h4}Ethene}

\begin{figure}[htp]
    \centering
    \includegraphics[width=1.0\linewidth]{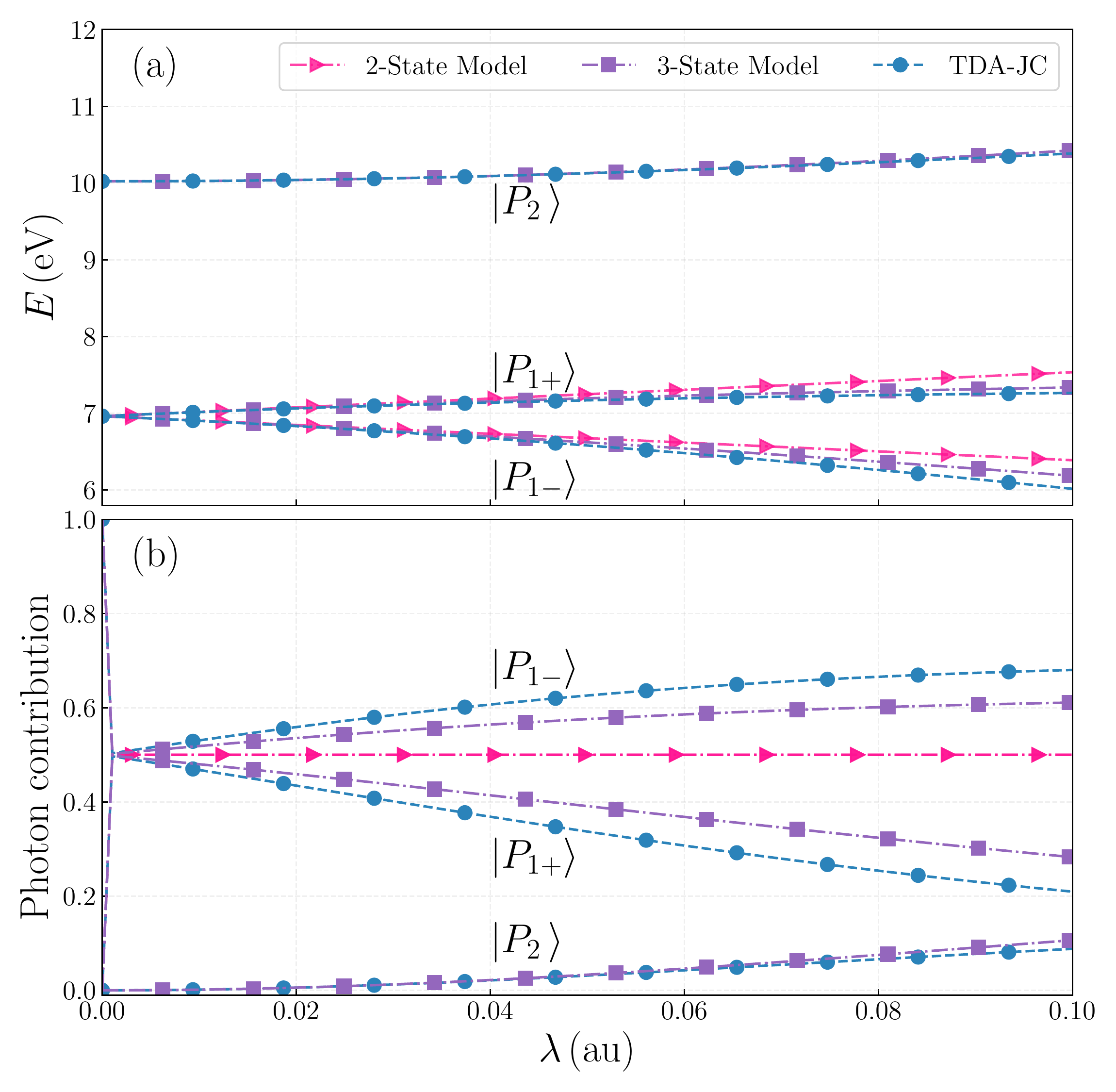}
    \caption{(a) Polariton spectrum of the ethene molecule and (b) photon contribution to each polariton state within the TDA-JC model using PBE0 functional and 6-311++G** basis set, as a function of the coupling strength $\lambda$. Solutions to model Hamiltonians with two or three states are also presented for a comparison. 
    }
    \label{fig::c2h4}
\end{figure}

For the ethene molecule, the first excited state in the gas-phase has an TDA excitation energy of $6.961 \, \mathrm{eV}$ and a transition dipole moment in the $x$--direction. 
As shown in Fig.~\ref{fig::c2h4}a, this state exhibits an expected Rabi splitting into two polariton states upon the application of a resonant radiation field in the $x$--direction. 
Within the TDA-JC model, the lower polariton, $\left| P_{1-} \right\rangle$, keeps reducing its energy while the upper polariton, $\left| P_{1+} \right\rangle$, is subjected to a monotonic increase in its energy.  
\textit{The Rabi splitting is clearly becoming non-symmetric.} By the maximum coupling strength, $\lambda = 0.1038 \, \mathrm{au}$, the lower polariton has a TDA-JC excitation energy of $5.965 \,  \mathrm{eV}$, which translated to a net reduction of $0.996 \,  \mathrm{eV}$ from its gas-phase value.  In contrast, the upper polariton is predicted to have an TDA-JC excitation energy value of $7.270 \, \mathrm{eV}$ at $\lambda = 0.1038 \, \mathrm{au}$, which amounts to a net gain of only $0.308 \, \mathrm{eV}$.

A symmetric Rabi splitting of the upper and lower polaritons would be expected for a 2-state Jaynes-Cummings model Hamiltonian, 
which is constructed using gas-phase TDA results according to Eq. \ref{eq:two-state-Hamiltonian}.
The energy eigenvalues, as expressed in Eqs. \ref{eq:Omega1-C} and \ref{eq:Omega1+C}, are plotted (as pink triangles) against the coupling strength in Fig.~\ref{fig::c2h4}a.  
The energy of the lower polariton within the 2-state model decreases linearly with the electron-photon coupling strength (up to a net reduction of $0.595 \, \mathrm{eV}$), while that of the upper polariton increases linearly (in a symmetric fashion with respect to the lower polariton).     
\textit{While the 2-state model can be considered as an appropriate approximation when the coupling strength is weak, a non-symmetric Rabi splitting at stronger coupling strength within the TDA-JC model would suggest a non-negligible perturbation from one or more higher excited states.}

Within our current implementation of QED-TDA and QED-TDDFT methods, all molecules are assumed to adopt the same orientation.  (An implementation for randomly oriented photochromes\cite{cwik_excitonic_2016} is under development and shall be presented in  subsequent publications.)  As a direct consequence of this assumption, among higher excited states, only those with a large transition dipole component in the $x$--direction can perturb the aforementioned polaritons.  As shown in Table S1, 
the next state with a substantial dipole moment in the $x$--direction is the 13-th state with an excitation energy of $10.022 \, \mathrm{eV}$.  This state (labelled as $\left| P_2 \right\rangle$ in Fig.~\ref{fig::c2h4}a) could be accounted for through building a 3-state Hamiltonian in Eq. \ref{eq:TDA-JC-3-state}.  For both $\left| P_{1-} \right\rangle$ and $\left| P_{1+} \right\rangle$ polariton states, their energies within the 3-state model are brought much closer to the TDA-JC values in Fig.~\ref{fig::c2h4}a, which is a substantial improvement over the 2-state model. 

Interestingly, upon the perturbation from the higher excited states, the $\left| P_{1-} \right\rangle$ and $\left| P_{1+} \right\rangle$ polariton states both get lowered in their energies, hence producing the non-symmetric Rabi splitting.  Such energy lowerings can be easily understood within the 3-state model, where both polaritons are shown in Eqs. \ref{eq:Omega1-C2} and \ref{eq:Omega1+C2} to receive an identical and negative second-order contribution to their energies.  To offset these energy changes, the energy of the $\left| P_2 \right\rangle$ state (corresponding to the 13-th excited state in the gas--phase) gains energy with increasing coupling strengths as demonstrated in Fig.~\ref{fig::c2h4}a.

As far as the polariton ``wavefunction'' is concerned, Fig.~\ref{fig::c2h4}b shows that, at the weak--coupling limit, $\left| P_{1-} \right\rangle$ and $\left| P_{1+} \right\rangle$ each contains a 50\% of photon contribution.  
As the coupling strength increases, however, the TDA-JC $\left| P_{1-} \right\rangle$ state (as well as the one from the 3-state model) gradually gains more photon character.  An explanation of this again requires us to go beyond the 2-state model, which incorrectly predicts a non-varying photon contribution.  Indeed, within the 3-state model, the lower-polariton ``wavefunction'' in Eq. \ref{eq:3-state-lower-wfn} contains larger and larger contributions from $\ket{g} \ket{1}$ with increasing coupling strengths. In contrast, the TDA-JC upper polariton,  $\ket{P_{1+}}$, as well as its 3-state counterpart gradually loses photon character and,  as a compensation, the $\ket{P_2}$ state slowly gains some photon character.

In terms of the photon character of $\ket{P_{1-}}$ and $\ket{P_{1+}}$ states, the 3--state model only qualitatively predicts the trend of the TDA-JC model.
To fully reproduce the TDA-JC energies {within $0.02 \, \mathrm{eV}$},  however, it would take at least additional 18 excited states from the gas-phase calculation in the construction of the model Hamiltonian.  
\textit{This reflects the strength of our QED-TDA (and QED-TDDFT) algorithms: instead of hand--picking excited states that might exert a significant perturbation, these states are automatically accounted for through the Davidson diagonalization procedure.}

Our TDA-JC calculations and 3-state modeling are carried out with only up to the maximum coupling strength that is allowable by the molecular volume.  Theoretically, though, if one goes beyond that limit, the 3-state model will show (a) the photon character of $\ket{P_{1-}}$ reaches a peak value before decreasing, and (b) the upper polariton, $\ket{P_{1+}}$, loses all its photon character and converges its energy to the value in Eq. \ref{eq:Omega1+-3-state-limit}.

\begin{figure*}[htp]
    \centering
    \includegraphics[width=0.95\linewidth]{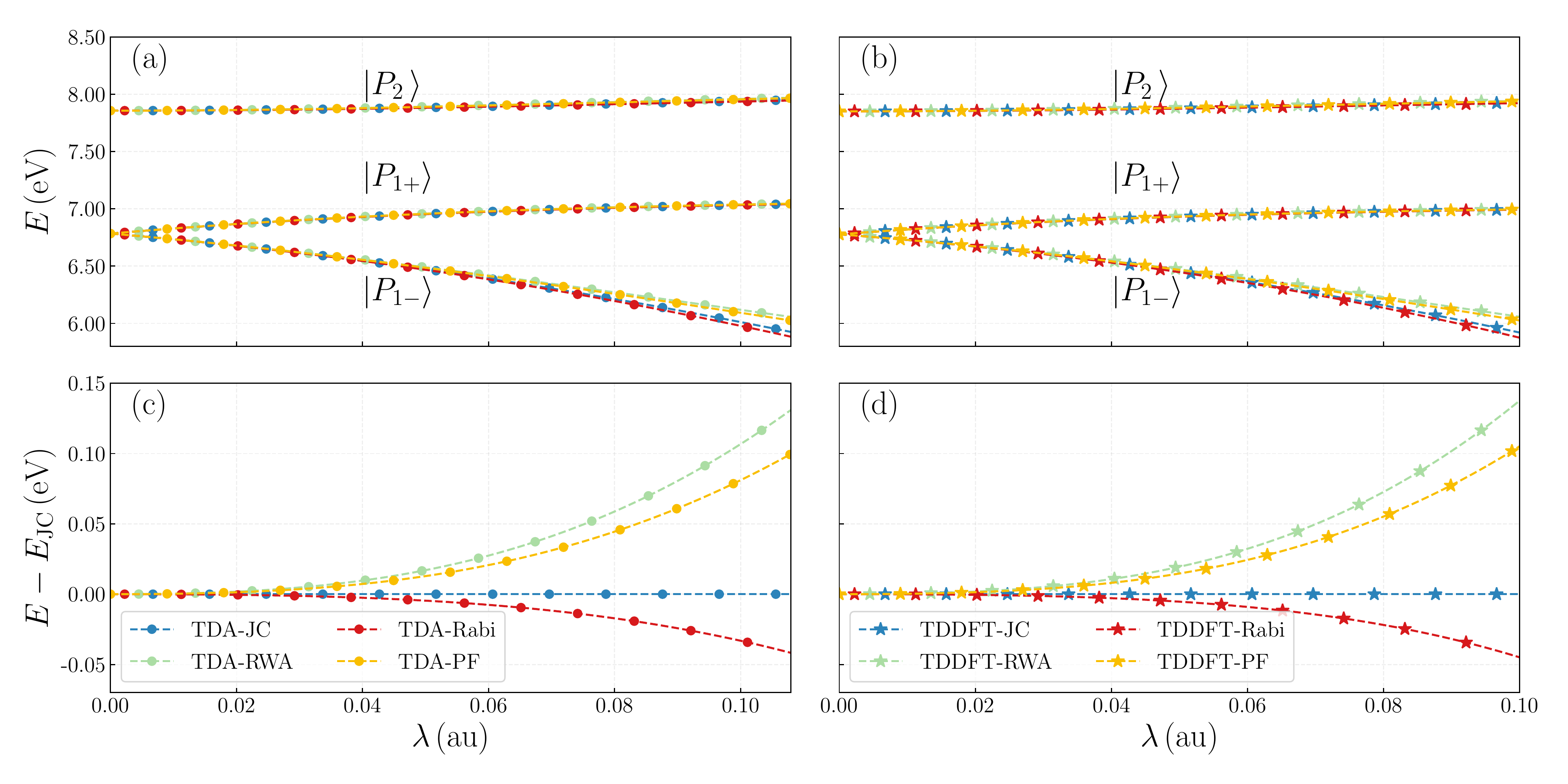}
    \caption{(a) QED-TDA and (b) QED-TDDFT polariton spectra of the formaldehyde molecule using different QED-TDDFT models with PBE0 functional and 6-311++G** basis set, as a function of the coupling strength $\lambda$.  Panels (c) and (d) highlight the difference in the lower polariton energy from Rabi, RWA and PF models against corresponding TDA-JC or TDDFT-JC values.  
    }
    \label{fig::ch2o}
\end{figure*}




\subsection{\label{sec::ch2o}Formaldehyde}

For the formaldehyde molecule, the TDA-JC model (marked as blue dots in Fig.~\ref{fig::ch2o}a) yielded very similar results to that of ethene: the first excited state with a substantial oscillator strength in the gas-phase (excitation energy: $6.784 \, \mathrm{eV}$; dipole: $0.490 \, \mathrm{au}$) undergoes a non-symmetric Rabi splitting in the photon cavity. This happens due to a perturbation from the 4-th state (labelled as $\ket{P_2}$) with an excitation energy of $7.856 \, \mathrm{eV}$ and a large transition dipole moment of $0.3728 \, \mathrm{au}$ in the $x$-direction.

For this molecule, we would shift our attention to a comparison among the  JC, Rabi, RWA, and PF variations of QED-TDA or QED-TDDFT methods. As shown in Fig~\ref{fig::ch2o}a, when the coupling strength is small, the four variations produce very similar energy spectra. In fact, at $\lambda = 0.045 \, \mathrm{au}$, the predicted Rabi splitting differ by no more than $0.02 \, \mathrm{eV}$ among the four models. 


When the coupling strength further increases, the four models still yield nearly identical energies for the upper polariton, $\ket{P_{1+}}$.  However, the predicted energies for the lower polariton, $\ket{P_{1-}}$, start to exhibit noticeable differences (Fig~\ref{fig::ch2o}a). This is more clear in Fig~\ref{fig::ch2o}c,  which shows the energy differences against the TDA-JC model.
The TDA-Rabi model (marked red), which adds the CRT term to TDA-JC, lowers the polariton energy in an agreement with the leading perturbative correction in Eq.~\ref{eq::crt}. 
Meanwhile, the TDA-RWA model (illustrated as green dots) captures the DSE contribution missing in the TDA-JC model and thus raises the polariton energy in consistence with Eq.\ref{eq::dse}.  At large $\lambda$ values, the CRT correction to TDA-JC model is found to be 3--4 times smaller than the DSE correction, in terms of the absolute value. For the TDA-PF model (marked yellow in Fig~\ref{fig::ch2o}b), which adds both CRT and DSE corrections to the TDA-JC model, the CRT term only partially cancelled the DSE component, also leading to a net energy increase.

\begin{figure*}[htp]
    \centering
    \includegraphics[width=1.0\linewidth]{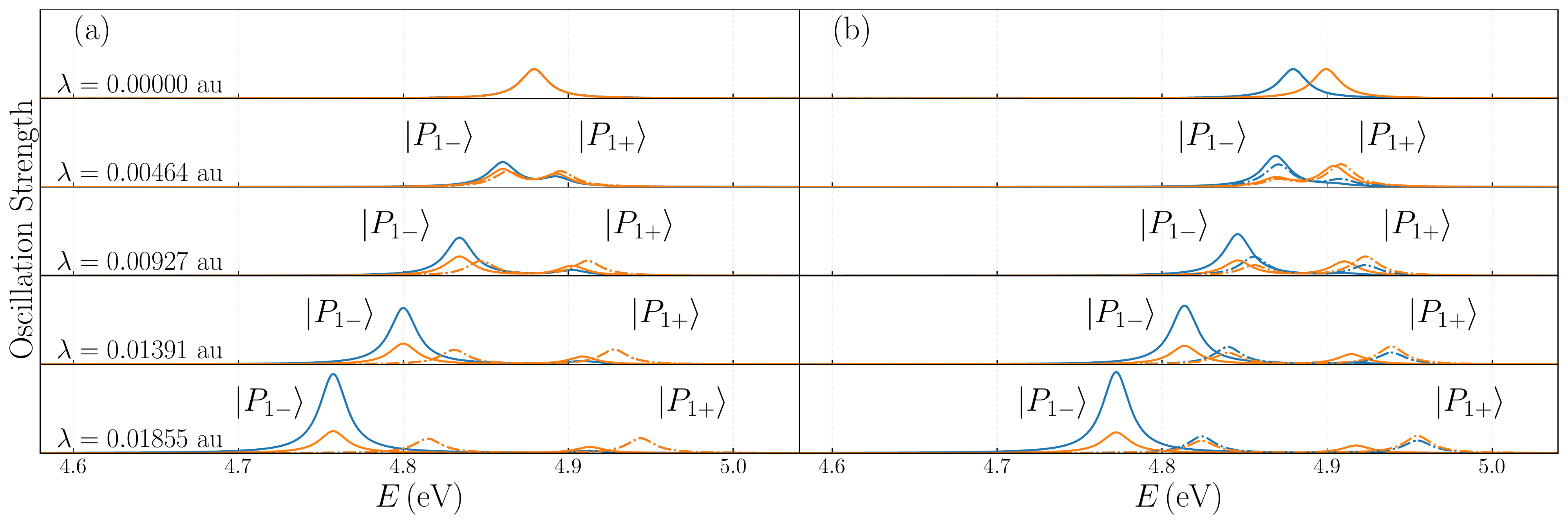}
    \caption{{Absorption} spectra of the benzaldehyde molecule from TDA-JC (solid lines) and 2-state Jaynes-Cummings model ({dot-}dashed lines) calculations {with different coupling strengthes} using the PBE0 functional and 6-311++G** basis set. Fabry-P\'{e}rot mode is chosen to be (a) in resonance ($4.879$ eV) and (b) 0.02 eV off-resonance ($4.899$ eV) from the gas-phase excited state. {The blue and orange lines indicate the electron and photon contributions, respectively.}  Following Ref. \citenum{flick_lightmatter_2019}, Lorentzian broadening is employed with $\Delta = 10^{-2} \, \mathrm{eV}$.}
    \label{fig::c6h5cho}
\end{figure*}

At smaller $\lambda$ values, the CRT correction is found numerically to be exactly $-\frac{1}{4}$ of the DSE correction, which is consistent with our earlier analysis at the end of Subsection~\ref{sec::prism}.  There we also mentioned that Huo and coworker's prediction that the CRT correction is $-\frac{1}{2}$ of the DSE value at resonance.\cite{mandal_polariton-mediated_2020}  This discrepancy is caused by a subtle difference in choosing the value of the fundamental coupling strength, $\bm{\lambda}_\alpha$, which is set to be $\sqrt{\frac{2}{\epsilon_0 V_\mathrm{eff}}} \boldsymbol{\epsilon}_\alpha$ in our work but $\sqrt{\frac{1}{\epsilon_0 V_\mathrm{eff}}} \boldsymbol{\epsilon}_\alpha$ in Ref. \citenum{mandal_polariton-mediated_2020}.  As a result, our DSE correction in Eq. \ref{eq::dse} is twice larger.  Meanwhile, at resonance and weak coupling, the electron-photon coupling in Eq. \ref{eq:g_alpha_I} can be written as $ g_{\alpha,I} = \sqrt{\frac{ \omega_\alpha}{2 \hbar}}  \left( \bm{\lambda}_\alpha \cdot \bm{\mu}_I^{\textrm{TDA-JC}} \right) = \sqrt{\frac{ \omega_\alpha}{2 \hbar}}  \left( \bm{\lambda}_\alpha \cdot \frac{1}{\sqrt{2}}\bm{\mu}_I^{\textrm{TDA}} \right) = \sqrt{\frac{ \omega_\alpha}{2 \hbar \epsilon_0 V_\mathrm{eff}}}  \left( \boldsymbol{\epsilon}_\alpha \cdot \bm{\mu}_I^{\textrm{TDA}} \right)$, which is exactly the same as Eq. 11 in Ref. \citenum{mandal_polariton-mediated_2020}.  This led to an identical CRT correction, $-\frac{\hbar}{2 \omega_\alpha}g_{\alpha,I}^2$.
In choosing our value for $\bm{\lambda}_\alpha$, we follow Rubio,\cite{ruggenthaler_quantum-electrodynamical_2014,flick_lightmatter_2019} Subotnik,\cite{chen_ehrenfestr_2019-1,chen_ehrenfestr_2019,li_quasiclassical_2020,li_cavity_2021} and others.\cite{tokatly_time-dependent_2013,deprince_cavity-modulated_2021} It would be a reasonable choice when one or more molecules is placed within a horizontal plane equidistant from the two mirrors ($z = L_z/2$, where the sine wave reaches its maximum value).  
When the molecular concentration further increases, however, it might be better to follow Huo and coworkers \cite{mandal_investigating_2019,mandal_polariton-mediated_2020} and adopt a reduced $\bm{\lambda}_\alpha$ value to reflect a vertical molecular distribution.

While our discussion on the formaldehyde molecule has so far focused on QED-TDA calculations, all our observations on the comparison among JC, Rabi, RWA, and PF models would also apply to QED-TDDFT results displayed in Fig.~\ref{fig::ch2o}c and ~\ref{fig::ch2o}d.
Overall, with the leading contribution from CRT and DSE being second order to the coupling strength, both terms can be ignored for small coupling strengths. Namely, \textit{the TDA-JC and TDDFT-JC models can be used to describe polariton states at  weak light-matter interaction regime.}\cite{stokes_gauge_2019,mandal_polariton-mediated_2020}  However, more caution is needed to select an appropriate model in strong and ultra-strong coupling regimes, where
the lower polariton energy can become unbounded without the DSE correction.\cite{mandal_polariton-mediated_2020}



\subsection{\label{sec::c6h5cho}Benzaldehyde}

The TDA-JC results of the benzaldehyde molecule are presented as a more realistic example of the photochrome.  (Other QED-TDDFT models are also tested and found to lead to similar results in Table S5.)
In particular, the focus is placed on the second excited state from the gas-phase TDA calculation with an excitation energy of $4.879 \, \mathrm{eV}$ and a transition dipole of $0.427 \, \mathrm{au}$ in the $xy$--plane.  
In Fig~\ref{fig::c6h5cho}a, this state is coupled to a resonant Fabry-P\'{e}rot mode; while in Fig~\ref{fig::c6h5cho}b, it is coupled to an off-resonance cavity mode of $4.899 \, \mathrm{eV}$ (\textit{i.e.}, a detuning of $0.02 \, \mathrm{eV}$).   
Clearly, as the coupling strength $\lambda$ increases, larger Rabi splittings occur in both resonant and off-resonant cases.  Moreover, the energies of both lower and upper polaritons deviate further away from the 2-state-predicted values (dot-dashed lines), in a way similar to the ethene and formaldehyde molecules. 
Interestingly, for our chosen coupling strengths, the off-resonance results in Fig.~\ref{fig::c6h5cho}b show slightly smaller differences between the 2-state Jaynes-Cummings model and the TDA-JC method.
This occurs because, with a detuning of 0.02 eV, the overall Rabi splitting is slightly smaller. 

\begin{figure}[htp]
    \centering
    \includegraphics[width=0.9\linewidth]{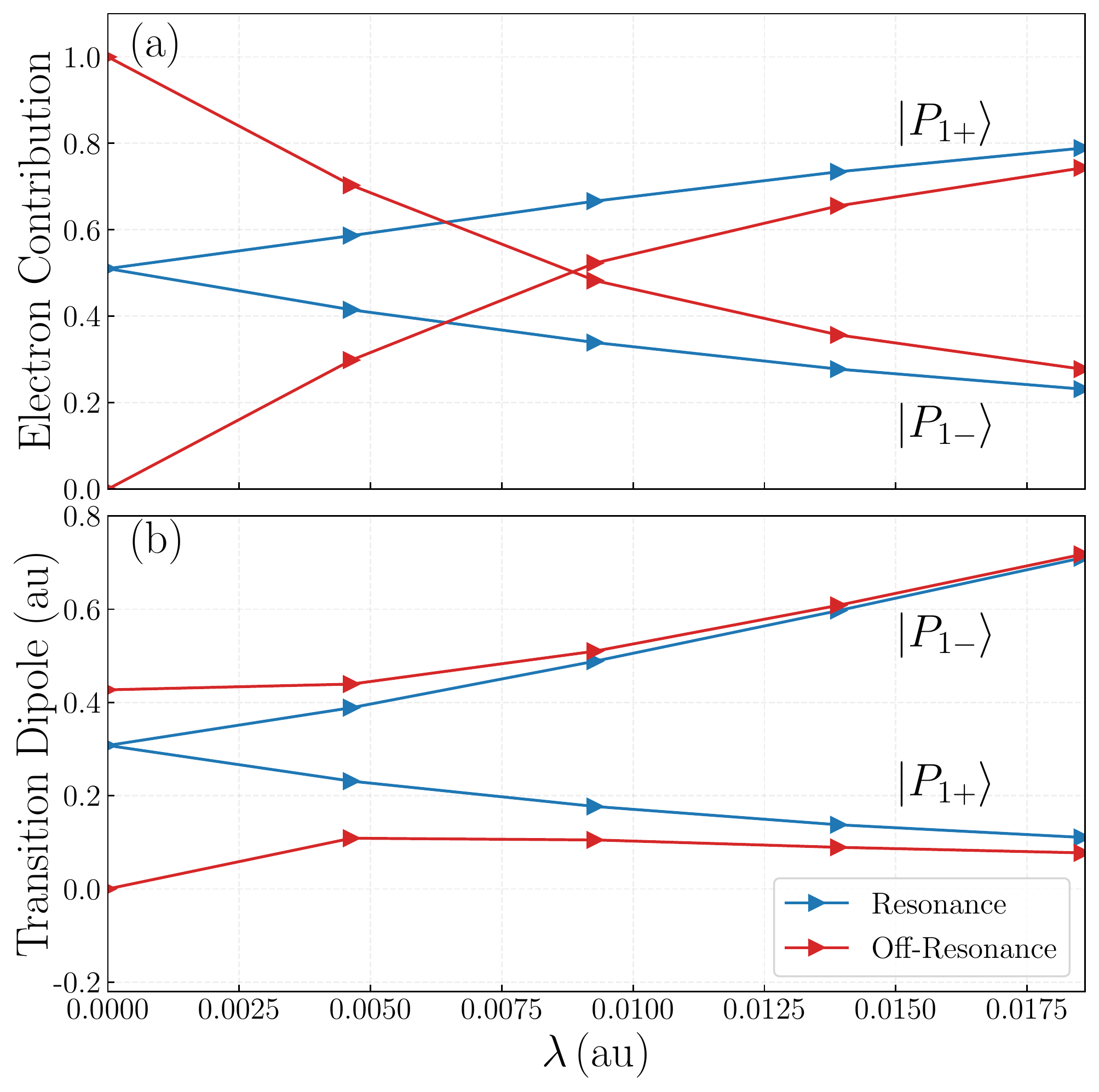}
    \caption{(a) Electron contributions {and (b) the corresponding norms of transition dipole moments for} the lower and upper polaritons {of bezaldehyde} at different coupling strengths {with resonance (blue) and off-resonance (red) photon energies.}}
    \label{fig::c6h5cho-2}
\end{figure}

An analysis of the composition of polariton ``wavefunction'' of benzaldehyde leads to observations similar to the ethene molecule (Fig. \ref{fig::c2h4}b).  At non-zero coupling strengths, the lower polariton, $\ket{P_{1-}}$, is shown in Fig. \ref{fig::c6h5cho-2}a to contain less electron contribution than the photon contribution, while the opposite can be seen for the upper polariton, $\ket{P_{1+}}$.
Such deviations from the 2-state Jaynes-Cummings model, which would predict equal electron and photon contributions for both polaritons in a resonance coupling, can be traced to non-negligible second-order contributions from higher excited states (3rd, 5th, and 11-th states in Table S4; and many more) of gas-phase benzaldehyde.  For off-resonance coupling, similar behavior can also be seen in Fig. \ref{fig::c6h5cho-2}b, but the lower polariton, $\ket{P_{1-}}$, contains more electron contributions (than the resonance case) due to a higher photon energy.  

At first glance, smaller electronic contribution to the lower polariton with stronger electron--photon coupling as shown in Fig. \ref{fig::c6h5cho-2}a would appear to contradict Fig.~\ref{fig::c6h5cho}a and \ref{fig::c6h5cho}b, where the TDA-JC molecular oscillation strength (marked blue) of the lower polariton actually \textit{increases} with the coupling strength.   
To resolve such a ``contradiction'', it is useful to examine the change of the molecular transition dipoles of the polaritons (as defined in Eq. S2 and S3 in the SI) with {varying} coupling {strength}.  Surprisingly, {as shown in Fig. \ref{fig::c6h5cho-2}b}, the transition dipole of the lower polariton increases with larger coupling strength, while the opposite happens to the upper polariton.  This can be understood within a 3--state model.  Specifically, Eq. \ref{eq:tm-lower} indicates that the transition dipole of the lower polariton will be enhanced by those of higher-energy excited states, while the transition dipole of the upper polariton will get weakened by those states (Eq. \ref{eq:tm-upper}).  
This is confirmed by Fig. S1, which shows a steady increase in the transition dipole of the lower polariton as more states are included the Jaynes-Cummings model Hamiltonian, and by an opposite trend for the upper polariton in the Figure.  
In the limit of strong couplings (such as $\lambda = 0.01853$ au), the upper polariton state of benzaldehyde should be dominated by the $\ket{e_1}$ state (with the corresponding coefficient being 0.88). But due to the slow $\frac{1}{\omega_2 - \omega}$ decay arising from Eqs. \ref{eq:tm-lower} and \ref{eq:tm-upper} and demonstrated in Fig. S1, many higher excited states (with small but nonzero mixing coefficients) combine together to cancel the $\ket{e_1}$ dipole moment.  At $\lambda = 0.01853$ au, the transition dipoles for the upper polariton is reduced by two thirds from its gas-phase value, leading to only marginal oscillator strengths in Fig.~\ref{fig::c6h5cho}.  

Such collective enhancement (weakening) of the lower (upper) polariton transition dipole by many higher excited states would be difficult to capture with the construction of multi-state Jaynes-Cummings model Hamiltonians. As shown in Fig. S1, with strong coupling, tens of excited states need to be included in the model before the converged value of the transition dipole moment can be approached.  But it holds the key to our understanding of the absorption spectra of benzaldehyde in this work as well as other molecules displaying non-symmetric Rabi splitting, such as merocyanine in Fig. 3a of Ref.\citenum{hutchison_modifying_2012}.

\section{Conclusions}
\label{sec::conclusions}

In summary, some progress has been made in this work on the formulation, implementation, and understanding of QED-TDDFT models, including,
\begin{itemize}
\item The 2-state Jaynes-Cummings model could be considered as an appropriate approximation when the coupling strength is weak enough that the higher excited states can be ignored.
\item Through linear-response and equation-of-motion formulations, simple QED-TDDFT working equations are obtained for the Pauli-Fierz Hamiltonian.  The Gaussian-basis implementation of the TDDFT-PF and associated approximate models paves the way for their routine applications.  
\item  {In the strong and ultra-strong coupling regime, the polaritons might get perturbed noticeably by higher excited states with significant transition dipole moments.} For our test molecules, these excited states reduce the electron contribution to the lower polariton but enhance its transition dipole moment and oscillator strength, whereas they affect the upper polariton in exactly the opposite manner.
While the QED-TDDFT/TDA method accounts for these effects naturally, it would be cumbersome (if feasible at all with the slow decay) to identify all these states from gas-phase calculations and then include them in the model Hamiltonian. 
\item At the strong coupling limit, the dipole self energy and counter rotation terms can also cause noticeable changes to the energies of polariton states. 
\end{itemize} 

On the other hand, within the TDDFT framework, several technical components have yet to be developed: 
\begin{itemize} 
\item In this work, the photochromes are assumed to be separated in the vacuum.  To capture the effect of other molecular species in the cavity (such as PMMA or solvents), implicit solvent models or combined quantum mechanical molecular mechanical models should be adopted. 
\item Analytical energy derivatives are needed to allow (a) the optimization of the photochoromic geometry for different polariton states and (b) the modeling of photochemical reactions and absorption/emission spectra.  
\item All molecules are assumed to adopt the same orientation within the optical cavity.  Our models need to be extended to cases where molecules adopt random orientations. 

\end{itemize} 
Work along these lines is expected to be rather straightforward and will be reported in subsequent publications. 

\begin{acknowledgments}
YS is supported by the National Institutes of Health (Grant No. R01GM135392), 
Oklahoma Center for the Advancement of Science and Technology (Grant No. HR18-130), 
and the Office of the Vice President of Research and the College of Art and Sciences at the University of Oklahoma (OU).
ZS acknowledges finial support from the National Natural Science Foundation of China (Grant No. 21788102) as well as the Ministry of Science and Technology of China through the National Key R\&D Plan (Grant No. 2017YFA0204501).
QO acknowledges financial support from the National Natural Science Foundation of China (Grant No. 22003030), China Postdoctoral Science Foundation (Grant No.2020M670280), and the Shuimu Tsinghua Scholar Program.  
All calculations were performed using computational resources at the OU Supercomputing Center for Education and Research (OSCER). 
\end{acknowledgments}

\section*{Data Availability} 
The data that support the findings of this study are available from one of the corresponding authors (YS) upon reasonable request.   The code implementation will be made available through {\sc PySCF} and {\sc Q-Chem}. 

\appendix

\section{Linear-Response Derivation}
In this section, we present a matrix derivation of the QED-TDDFT equations, which closely follows Rubio and coworkers' approach based on linear-response formula.\cite{tokatly_time-dependent_2013,ruggenthaler_quantum-electrodynamical_2014,pellegrini_optimized_2015,flick_kohnsham_2015,flick_cavity_2017,flick_atoms_2017,tokatly_conserving_2018,flick_ab_2018}
\label{appendix::lr}
\subsection{Density Response Kernel} 
For a KS-DFT electronic ground-state, its Fock matrix (\textit{i.e.} effective one-electron Hamiltonian) is diagonal in the representation of Kohn-Sham orbitals, 
\begin{eqnarray}
\mathbf{F}_0 = 
\begin{bmatrix} \mathbf{F}_{\textrm{oo}} & \mathbf{0} \\ \mathbf{0} & \mathbf{F}_{\textrm{vv}} \end{bmatrix} 
\end{eqnarray}
with $F_{ij} = \varepsilon_i \delta_{ij}$ and $F_{ab} = \varepsilon_a \delta_{ab}$.  The density matrix takes the format, 
\begin{eqnarray}
\mathbf{P}_0 = 
\begin{bmatrix} \mathbf{I}_{\textrm{oo}} & \mathbf{0} \\ \mathbf{0} & \mathbf{0} \end{bmatrix} 
\end{eqnarray}
with a single occupancy for the lowest-energy spin orbitals. 

Within the matrix formulation of TDDFT, the Fock matrix is subjected to a frequency-dependent perturbation, 
\begin{eqnarray}
\mathbf{F}(t) & = & \mathbf{F}_0 + \delta \mathbf{F}(t)  \\
\delta \mathbf{F}(t) & =  & 
\begin{bmatrix} \mathbf{0} & \delta\mathbf{F}_{\textrm{ov}}^{(\Omega)} e^{i\Omega t}     
\\  \delta\mathbf{F}_{\textrm{vo}}^{(\Omega)}  e^{-i\Omega t}   & \mathbf{0} \end{bmatrix} \nonumber \\
&  & +\begin{bmatrix} \mathbf{0} &  \delta\mathbf{F}_{\textrm{ov}}^{(-\Omega)} e^{-i\Omega t}  
\\ \delta\mathbf{F}_{\textrm{vo}}^{(-\Omega)}  e^{i\Omega t}   & \mathbf{0} \end{bmatrix} 
\end{eqnarray}
where $\delta\mathbf{F}_{\textrm{ov}}^{(\Omega)} = \delta\mathbf{F}_{\textrm{vo}}^{(\Omega),\dagger}$ and 
$\delta\mathbf{F}_{\textrm{ov}}^{(-\Omega)} = \delta\mathbf{F}_{\textrm{vo}}^{(-\Omega),\dagger}$ to maintain a Hermitian matrix.  Such a perturbation causes a response in the density matrix, 
\begin{eqnarray}
\mathbf{P}(t) & = & \mathbf{P}_0 + \delta \mathbf{P}(t)  \\
\delta \mathbf{P}(t) & =  & 
\begin{bmatrix} \mathbf{0} & \delta\mathbf{P}_{\textrm{ov}}^{(\Omega)} e^{i\Omega t}     
\\  \delta\mathbf{P}_{\textrm{vo}}^{(\Omega)}  e^{-i\Omega t}   & \mathbf{0} \end{bmatrix} \nonumber \\
&  & +\begin{bmatrix} \mathbf{0} &  \delta\mathbf{P}_{\textrm{ov}}^{(-\Omega)} e^{-i\Omega t}  
\\ \delta\mathbf{P}_{\textrm{vo}}^{(-\Omega)}  e^{i\Omega t}   & \mathbf{0} \end{bmatrix}   \label{eq:deltaP}
\end{eqnarray}
where $\delta\mathbf{P}_{\textrm{ov}}^{(\Omega)} = \delta\mathbf{P}_{\textrm{vo}}^{(\Omega),\dagger}$ and 
$\delta\mathbf{P}_{\textrm{ov}}^{(-\Omega)} = \delta\mathbf{P}_{\textrm{vo}}^{(-\Omega),\dagger}$.

The density response is governed by the time-dependent Kohn-Sham (TDKS) equation,\cite{casida_time-dependent_1995,hirata_time-dependent_1999,dreuw_single-reference_2005,chen_random-phase_2017}
\begin{align}
i\hbar \frac{\partial \mathbf{P}(t)}{\partial t} & =  & [\mathbf{F}(t), \mathbf{P}(t)]  = [\delta \mathbf{F}(t), \mathbf{P}_0] + [\mathbf{F}_0, \delta \mathbf{P}(t)]
\end{align} 
which is  
\begin{align} 
& i\hbar \frac{\partial \mathbf{P}(t)}{\partial t} 
 = \begin{bmatrix} \mathbf{0} & -\delta\mathbf{F}_{\textrm{ov}}  \\  \delta\mathbf{F}_{\textrm{vo}}   & \mathbf{0} \end{bmatrix}  \nonumber \\
& + \begin{bmatrix} \mathbf{0} & \mathbf{F}_{\textrm{oo}} \delta\mathbf{P}_{\textrm{ov}} - \delta\mathbf{P}_{\textrm{ov}} \mathbf{F}_{\textrm{vv}} 
\\ \mathbf{F}_{\textrm{vv}}  \delta\mathbf{P}_{\textrm{vo}} - \delta\mathbf{P}_{\textrm{vo}}  \mathbf{F}_{\textrm{oo}}  & \mathbf{0} \end{bmatrix} 
\end{align}

Below we shall focus on the vo-block of this equation, which requires \begin{equation}
 i\hbar \frac{\partial \mathbf{P}_{\textrm{vo}}(t)}{\partial t} = \delta\mathbf{F}_{\textrm{vo}} + 
  \mathbf{F}_{\textrm{vv}}  \delta\mathbf{P}_{\textrm{vo}} - \delta\mathbf{P}_{\textrm{vo}}  \mathbf{F}_{\textrm{oo}},  \label{eq:tdks-2} 
\end{equation} 
noting that the ov-block equation is the complex conjugate.  

Collecting the $e^{-i\Omega t}$ and $e^{i\Omega t}$ terms separately from both sides of Eq. \ref{eq:tdks-2}, one gets
\begin{eqnarray}
(\hbar \Omega)  \delta\mathbf{P}_{\textrm{vo}}^{(\Omega)} 
& = &  \delta\mathbf{F}_{\textrm{vo}}^{(\Omega)}  + \mathbf{F}_{\textrm{vv}}  \delta\mathbf{P}_{\textrm{vo}}^{(\Omega)} 
- \delta\mathbf{P}_{\textrm{vo}}^{(\Omega)}  \mathbf{F}_{\textrm{oo}}  \\
(- \hbar \Omega) \delta\mathbf{P}_{\textrm{vo}}^{(-\Omega)} 
& = & \delta\mathbf{F}_{\textrm{vo}}^{(-\Omega)} + \mathbf{F}_{\textrm{vv}}  \delta\mathbf{P}_{\textrm{vo}}^{(-\Omega)} 
- \delta\mathbf{P}_{\textrm{vo}}^{(-\Omega)}  \mathbf{F}_{\textrm{oo}}  \nonumber \\
\end{eqnarray}
which can be written explicitly as 
\begin{eqnarray}
(\hbar \Omega  - \varepsilon_a + \varepsilon_i)  \delta P_{ai}^{(\Omega)}   & = &  \delta F_{ai}^{(\Omega)}   \label{eq:dm-response-1} \\
(-\hbar \Omega  - \varepsilon_a + \varepsilon_i)  \delta P_{ai}^{(-\Omega)} & = & \delta F_{ai}^{(-\Omega)}   \label{eq:dm-response-2}
\end{eqnarray}
This shows how the density matrix would respond to a frequency-dependent change in the Fock matrix. 

\subsection{Electron Equations within QED-TDDFT}

For a molecule in an optical cavity, its Fock matrix is influenced by changes in both the electronic density matrix in Eq. \ref{eq:deltaP} as well as the electron-photon coupling, 
\begin{eqnarray}
\delta F_{ai}^{(\Omega)} & = &  
\delta F_{ai,\textrm{elec}}^{(\Omega)} + 
\delta F_{ai,\textrm{elec-photon}}^{(\Omega)} \\
\delta F_{ai}^{(-\Omega)} & = &  
\delta F_{ai,\textrm{elec}}^{(-\Omega)} + 
\delta F_{ai,\textrm{elec-photon}}^{(-\Omega)} 
\end{eqnarray}
For $F_{ai}^{(\Omega)}$, it is only perturbed by $P_{bj}^{(\Omega)}$ and $P_{jb}^{(-\Omega)}$, 
which carry the same $e^{-i\Omega t}$ factor in Eq. \ref{eq:deltaP}, 
\begin{align}
& \delta F_{ai,\textrm{elec}}^{(\Omega)} =  \sum_{bj} 
\left( \frac { \partial F_{ai}} { \partial P_{bj}} \delta P_{bj}^{(\Omega)} 
+ \frac { \partial F_{ai}} { \partial P_{jb}} \delta P_{jb}^{(-\Omega)}  \right) \nonumber \\
& =  \sum_{bj} 
\left[ \left[ A_{ai,bj} - ( \varepsilon_a - \varepsilon_i) \delta_{ab} \delta_{ij} \right] \delta P_{bj}^{(\Omega)} +  B_{ai,bj}  \delta P_{jb}^{(-\Omega)}  \right] \label{eq:Faie-1} 
\end{align} 
Similarly, for $F_{ai,\textrm{elec}}^{(-\Omega)}$, one gets  
\begin{align}
& \delta F_{ai,\textrm{elec}}^{(-\Omega)}   =  \sum_{bj} 
\left( \frac { \partial F_{ai}} { \partial P_{bj}} \delta P_{bj}^{(-\Omega)} 
+ \frac { \partial F_{ai}} { \partial P_{jb}} \delta P_{jb}^{(\Omega)}  \right) \nonumber  \\
& =  \sum_{bj} 
\left[ \left[ A_{ai,bj} - ( \varepsilon_a - \varepsilon_i) \delta_{ab} \delta_{ij} \right] \delta P_{bj}^{(-\Omega)}   + B_{ai,bj}  \delta P_{jb}^{(\Omega)}  \right] 
\label{eq:Faie-2} 
\end{align}

For the electron-photon interaction energy in Eq. \ref{eq:Pauli-Fierz},
\begin{eqnarray}
V_{\textrm{elec-photon}} =   \sum_{\alpha=1}^M 
 \frac{1}{2} \omega_\alpha^2 \left( q_\alpha - \frac{ 1} { \omega_\alpha}  \bm{\lambda}_\alpha \cdot \bm{\mu} \right)^2 
\end{eqnarray}
its corresponding Fock matrix contribution is
\begin{align}
& F_{ai,\textrm{elec-photon}} 
= \frac{\partial V_{\textrm{elec-photon}}} 
{\partial P_{ai}} \nonumber \\  
& =  \sum_{\alpha=1}^M 
\omega_\alpha  \lambda_{ai}^\alpha
\left( q_\alpha - \frac{ 1} { \omega_\alpha}  \bm{\lambda}_\alpha \cdot \bm{\mu} \right) 
\label{eq:Fai-ep}
\end{align}
which uses the dipole derivative 
\begin{eqnarray}
\frac{\partial \bm{\mu}}{\partial P_{ai}} 
= \frac {\partial \left( \bm{\mu}_{\textrm{nuc}} - \bm{\mu}_{\textrm{mo}} \cdot \mathbf{P} \right) }
{\partial P_{ai}} = - \bm{\mu}_{ai}
\end{eqnarray}
Clearly, the Fock matrix contribution in Eq. \ref{eq:Fai-ep} is perturbed by the density matrix and photon coordinate, 
\begin{align}
& \delta F_{ai,\textrm{elec-photon}}^{(\Omega)}  =  \delta F_{ai,\textrm{elec-photon}}^{(-\Omega)} \nonumber \\
& =  \sum_{bj}  \left( \sum_{\alpha=1}^M  \lambda_{ai}^\alpha \lambda_{bj}^\alpha 
\right) 
\left( \delta P_{bj}^{(\Omega)} + \delta P_{jb}^{(-\Omega)} \right)  
+ \sum_{\alpha=1}^M \omega_\alpha  \lambda_{ai}^\alpha  \delta q_\alpha \nonumber \\
& = \sum_{bj}  \Delta_{ai,bj}  \left( \delta P_{bj}^{(\Omega)} + \delta P_{jb}^{(-\Omega)} \right)  \nonumber \\
& \quad + \sum_{\alpha=1}^M   \sqrt{ \frac {\hbar \omega_\alpha } {2} }    \lambda_{ai}^\alpha   \left( M_\alpha + N_\alpha \right) 
\label{eq:dFai-ep} 
\end{align}
which uses the expression of $\Delta_{ai,bj}$ in Eq. \ref{eq:delta-aibj} and replaces $\delta \hat{q}_\alpha$ with dimensional quantities, 
\begin{equation}
\delta q_\alpha  =  \sqrt{ \frac {\hbar} {2 \omega_\alpha } } \left(  M_\alpha + N_\alpha \right)  \label{eq:dqa-split}    
\end{equation}
Plugging Eqs. \ref{eq:Faie-1}, \ref{eq:Faie-2} and \ref{eq:dFai-ep} into the right-hand-side of Eqs. \ref{eq:dm-response-1} and \ref{eq:dm-response-2}, one obtains 
\begin{align}
& (\hbar \Omega)  \delta P_{ai}^{(\Omega)} =  
 \sum_{bj} \left( A_{ai,bj} + \Delta_{ai,bj} \right) \delta P_{bj}^{(\Omega)} \nonumber \\
 & \quad + \sum_{bj} \left( B_{ai,bj} + \Delta_{ai,bj} \right)  \delta P_{jb}^{(-\Omega)} \nonumber \\
& \quad + \sum_{\alpha=1}^M   \sqrt{ \frac {\hbar \omega_\alpha } {2} }   \lambda_{ai}^\alpha   \left( M_\alpha + N_\alpha \right) \\
& (-\hbar \Omega )  \delta P_{ai}^{(-\Omega)} =  
 \sum_{bj} 
 \left( A_{ai,bj} + \Delta_{ai,bj} \right) \delta P_{bj}^{(-\Omega)} \nonumber \\
& \quad + \sum_{bj} \left( B_{ai,bj} + \Delta_{ai,bj} \right)  \delta P_{jb}^{(\Omega)} \nonumber \\
& \quad + \sum_{\alpha=1}^M   \sqrt{ \frac {\hbar \omega_\alpha } {2} }   \lambda_{ai}^\alpha   \left( M_\alpha + N_\alpha \right) 
\end{align}
Assume that $\delta P_{ai}^{(\Omega)} = \delta P_{ia}^{(\Omega)}$ and $\delta P_{ai}^{(-\Omega)}=\delta P_{ia}^{(-\Omega)}$ are all real, and write them as $X_{ai}$ and $Y_{ai}$, respectively, we obtain the electronic portion of the QED-TDDFT equation in Eq. \ref{eq:tddft-pf}.

\subsection{Photon Equations within QED-TDDFT}

The photon equation of state is\cite{flick_lightmatter_2019}
\begin{align}
& \left( \frac{ \partial^2} {\partial t^2} + \omega_\alpha^2 \right) q_\alpha(t) 
 =  -\frac{1}{ \omega_\alpha}  j_\alpha^{\textrm{eff}} (t)\nonumber \\
& \quad \quad  =  -\frac{1}{ \omega_\alpha}  \left( j_\alpha (t) - \omega_\alpha^2  \bm{\lambda}_\alpha \cdot \bm{\mu}  \right)  \label{eq:photon-eos}
\end{align}
For $q_\alpha(t)$ and $j_\alpha^{\textrm{eff}} (t)$ oscillating with frequency $\Omega$, they satisfy 
\begin{eqnarray}
\left( \Omega^2 - \omega_\alpha^2 \right) \delta q_\alpha  =   \frac{1}{\omega_\alpha} \delta j_\alpha^{\textrm{eff}}
\end{eqnarray}
Therefore, the photon response is 
\begin{eqnarray}
 \delta q_\alpha  & = &    \frac{1}{\omega_\alpha}  \frac{1}{\Omega^2 - \omega_\alpha^2} \delta j_\alpha^{\textrm{eff}}  \nonumber \\
& = &  \frac{1}{2\omega_\alpha^2} \left( \frac{1}{\Omega - \omega_\alpha} - \frac{1}{\Omega + \omega_\alpha} \right) \delta j_\alpha^{\textrm{eff}}
\end{eqnarray}
A key to the QED-TDDFT method is the splitting of the right-hand-side,\cite{flick_lightmatter_2019} which underlies Eq. \ref{eq:dqa-split},
\begin{eqnarray}
\left( \hbar \Omega - \hbar \omega_\alpha \right) M_\alpha  = \sqrt { \frac{\hbar }{2 \omega_\alpha^3}}  \delta j_\alpha^{\textrm{eff}}  \\
- \left( \hbar \Omega + \hbar \omega_\alpha \right) N_\alpha = \sqrt { \frac{\hbar} {2 \omega_\alpha^3}}  \delta j_\alpha^{\textrm{eff}}
\end{eqnarray}
where $\delta j_\alpha^{\textrm{eff}}$ can be easily derived from Eq. \ref{eq:photon-eos}, 
\begin{eqnarray}
\delta j_\alpha^{\textrm{eff}, (\Omega)} & =  &  \omega_\alpha^2 
\sum_{bj} \lambda_{bj}^\alpha \left( \delta P_{bj}^{(\Omega)} + \delta P_{jb}^{(-\Omega)} \right)
\end{eqnarray}
Therefore, 
\begin{align}
(\hbar \Omega)   M_\alpha   =   \sum_{bj}  \sqrt{\frac{\hbar \omega_\alpha}{2}}  \lambda_{bj}^\alpha \left( \delta P_{bj}^{(\Omega)} + \delta P_{jb}^{(-\Omega)} \right) + (\hbar \omega_\alpha) M_\alpha \\
-(\hbar \Omega)  N_\alpha   =   \sum_{bj}
\sqrt{\frac{\hbar \omega_\alpha}{2}} \lambda_{bj}^\alpha \left( \delta P_{bj}^{(\Omega)} + \delta P_{jb}^{(-\Omega)} \right) + (\hbar \omega_\alpha)  N_\alpha 
\end{align}
which are the photon portion of Eq. \ref{eq:tddft-pf}.

\section{Equation-of-Motion Derivation}
\label{appendix::eom}
Below, we shall use an alternative approach, which based on the equation-of-motion by transforming into a Heisenberg picture, to derive the same QED-TDDFT equation in Eq. \ref{eq:tddft-pf}. \cite{scuseria_particle-particle_2013,ziegler_derivation_2014,yang_analysis_2020} Within this derivation, the split of $q_\alpha$ in Eq. \ref{eq:dqa-split} will come naturally. 
\subsection{Unitary Transformation of Electronic Wavefunction}

The electronic wavefunction evolves with an unitary transformation
\begin{eqnarray} 
\left| \Phi (t) \right> &  = &   e^{- \hat{\Lambda}_{\textrm{e}}(t)}  \left| \Phi_0 \right> \\
\hat{\Lambda}_{\textrm{e}}(t)  & = & \sum_{ai} \left( - \Theta_{ai}^\ast (t) \hat{a}_i^\dagger \hat{a}_a + \Theta_{ai} (t) \hat{a}_a^\dagger \hat{a}_i \right)
\end{eqnarray} 
Using the following commutators,\cite{scuseria_particle-particle_2013} 
\begin{eqnarray}
\left[\hat{a}_a^\dagger \hat{a}_i, \hat{a}_j^\dagger \hat{a}_b \right]  & =  & \delta_{ij} \hat{a}_a^\dagger \hat{a}_b - \delta_{ab} \hat{a}_j^\dagger \hat{a}_i  \\
\left[\hat{a}_a^\dagger \hat{a}_i, \hat{a}_b^\dagger \hat{a}_j \right]  & = & \left[ \hat{a}_i^\dagger \hat{a}_a, \hat{a}_j^\dagger \hat{a}_b \right]  =  0 
\end{eqnarray} 
one can find 
\begin{eqnarray}
\left[ \hat{\Lambda}_{\textrm{e}}, \hat{a}_j^\dagger \hat{a}_b \right] 
& = &   \sum_{a}  \Theta_{aj} (t) \hat{a}_a^\dagger \hat{a}_b - \sum_i  \Theta_{bi} (t)  \hat{a}_j^\dagger \hat{a}_i  \\
\left[ \hat{\Lambda}_{\textrm{e}}, \hat{a}_b^\dagger \hat{a}_j \right] 
& = &   \sum_{a}  \Theta_{aj}^*(t)  \hat{a}_b^\dagger \hat{a}_a - \sum_i  \Theta_{bi}^* (t)  \hat{a}_i^\dagger \hat{a}_j
\end{eqnarray} 
Using the leading terms of the following BCH expansion, 
\begin{align}
& e^{\hat{\Lambda}_{\textrm{e}} }  \left( \hat{a}_j^\dagger \hat{a}_b+\hat{a}_b^\dagger \hat{a}_j  \right)  e^{-\hat{\Lambda}_{\textrm{e}} }  
 =   \hat{a}_j^\dagger \hat{a}_b+\hat{a}_b^\dagger \hat{a}_j  \nonumber \\
&    +  \sum_{a} \left( \Theta_{aj} (t) \hat{a}_a^\dagger \hat{a}_b  +  \Theta_{aj}^*(t)  \hat{a}_b^\dagger \hat{a}_a  \right) \nonumber \\
&     - \sum_i \left( \Theta_{bi} (t)  \hat{a}_j^\dagger \hat{a}_i  + \Theta_{bi}^* (t)  \hat{a}_i^\dagger \hat{a}_j \right)  + \mathcal{O} (\Theta^2)    \label{eq:u-jb-ut} 
\end{align} 
one finds 
\begin{align}
& \left< \Phi (t) \middle| \left( \hat{a}_j^\dagger \hat{a}_b+\hat{a}_b^\dagger \hat{a}_j  \right)  \middle| \Phi (t) \right> \nonumber \\
& = \left< \Phi_0 \middle|   e^{\hat{\Lambda}_{\textrm{e}} }  \left( \hat{a}_j^\dagger \hat{a}_b+\hat{a}_b^\dagger \hat{a}_j  \right)  e^{-\hat{\Lambda}_{\textrm{e}} }   \middle| \Phi_0 \right> 
 =  - \Theta_{bj} (t) - \Theta_{bj}^* (t)   
 \label{eq:psi0-u-jb-ut-psi0}
\end{align} 

\subsection{Unitary Transformation of Photon Wavefunction}

Similarly, the photon wavefunction is also subjected to an unitary transformation (Eq.~26 in Ref.~\citenum{haugland_coupled_2020})
\begin{eqnarray}
\left| \chi (t) \right> &  = &  e^{- \hat{\Lambda}_{\textrm{ph}}(t)}  \left| \Phi_0 \right> \left| \chi_0 \right>  \\
\hat{\Lambda}_{\textrm{ph}} (t) & = & \sum_{\alpha} \left(-  C_{\alpha}^* (t) \hat{b}_{\alpha} + C_{\alpha} (t) \hat{b}_{\alpha}^\dagger \right)
\end{eqnarray}
Using the commutators for bosons, 
\begin{eqnarray}
\left[ \hat{b}_{\alpha}, \hat{b}_{\beta}^\dagger \right] = \delta_{\alpha \beta}, \quad   
\left[ \hat{b}_{\alpha}, \hat{b}_{\beta} \right]  = \left[ \hat{b}_{\alpha}^\dagger, \hat{b}_{\beta}^\dagger \right] = 0 
\end{eqnarray}
one gets
\begin{align}
\left[  \hat{\Lambda}_{\textrm{ph}},    \hat{b}_{\alpha} \right]  
 =   - C_{\alpha} (t),  \quad \quad 
\left[  \hat{\Lambda}_{\textrm{ph}},    \hat{b}_{\alpha}^\dagger \right]  
 =   - C_{\alpha}^* (t)  
\end{align}
Accordingly, the following BCH expansions vanish after the first order
\begin{eqnarray}
e^{\hat{\Lambda}_{\textrm{ph}}}   \hat{b}_{\alpha} e^{-\hat{\Lambda}_{\textrm{ph}}}  
= \hat{b}_{\alpha}  - C_{\alpha} (t) \\
e^{\hat{\Lambda}_{\textrm{ph}}}   \hat{b}_{\alpha}^\dagger e^{-\hat{\Lambda}_{\textrm{ph}}}  
=  \hat{b}_{\alpha}^\dagger  - C_{\alpha}^* (t) 
\end{eqnarray}
From this, one can compute 
\begin{align}
& \left< \chi (t) \middle| \left(\hat{b}_\alpha + \hat{b}_\alpha^\dagger  \right) \middle| \chi (t) \right> 
= - C_\alpha(t) - C_\alpha^*(t)  \label{eq:b_and_b_dagger}  \\
& \left< \chi (t) \middle|  \hat{b}_\alpha^\dagger  \hat{b}_\alpha  \middle| \chi (t) \right> \nonumber \\
& = \left< \chi_0 \middle|\left(  e^{\hat{\Lambda}_{\textrm{ph}}} \hat{b}_\alpha^\dagger e^{-\hat{\Lambda}_{\textrm{ph}}} \right)  \left( e^{\hat{\Lambda}_{\textrm{ph}}} \hat{b}_\alpha e^{-\hat{\Lambda}_{\textrm{ph}}} \right) \middle| \chi_0 \right>
= C_\alpha(t) C_\alpha^*(t)   \label{eq:b_dagger_b} 
\end{align} 

\subsection{Pauli-Fierz Energy Components} 

In the Pauli-Fierz Hamiltonian in Eq. \ref{eq:Pauli-Fierz}, the dipole moment operator can be written as 
\begin{align}
\hat{\bm{\mu}} =  \bm{\mu}_0  +  \Delta   \hat{\bm{\mu}} = \bm{\mu}_0  - \sum_{bj} \bm{\mu}_{bj} \left( \hat{a}_j^\dagger \hat{a}_b +  \hat{a}_b^\dagger \hat{a}_j \right)  \\
\bm{\lambda}_\alpha \cdot \hat{\bm{\mu}}  = \bm{\lambda}_\alpha \cdot \bm{\mu}_0 - \sum_{bj} \lambda_{bj}^\alpha \left( \hat{a}_j^\dagger \hat{a}_b +  \hat{a}_b^\dagger \hat{a}_j \right)   \label{eq:lambda-and-mu}
\end{align} 
where the ground-state dipole moment shifts the minimum of the quantum harmonic oscillator of the photon, $\hat{q}_\alpha \longrightarrow \hat{q}_\alpha -  \frac{ 1} { \omega_\alpha}  \bm{\lambda}_\alpha \cdot \bm{\mu}_0$.  Then, the Pauli-Fierz Hamiltonian can be written as 
\begin{eqnarray}
\hat{H}_{\textrm{PF}} & =  &  \hat{H}_\mathrm{elec}(t)  
+ \frac{1}{2} \sum_{\alpha=1}^M \left( \bm{\lambda}_\alpha \cdot \left< \Delta \hat{\bm{\mu}} \right> \right)^2 \nonumber \\
&  & - \sum_{\alpha=1}^M \omega_\alpha \hat{q}_\alpha \left( \bm{\lambda}_\alpha \cdot \left< \Delta \hat{\bm{\mu}} \right> \right)
\nonumber \\ & & + \sum_{\alpha=1}^M \left[ \frac{1}{2} \hat{p}_{\alpha}^2
+  \frac{1}{2} \omega_\alpha^2 \hat{q}_\alpha^2  \right]  
+ \sum_{\alpha=1}^M \frac{ j_\alpha(t) } { \omega_\alpha} \hat{q}_\alpha \quad \quad \quad \label{eq:Pauli-Fierz-2} 
\end{eqnarray} 

With this Hamiltonian, the energy of the time-evolving wavefunction is 
\begin{eqnarray}
E_{\textrm{PF}} (\bm{\Theta}, \mathbf{C})  = \left< \chi (t) \middle| \left< \Phi (t) \middle|   
\hat{H}_{\textrm{PF}} \middle| \chi (t) \right> \middle| \chi (t) \right> 
\end{eqnarray}
Its electronic component is\cite{ziegler_derivation_2014,scuseria_particle-particle_2013,yang_analysis_2020} 
\begin{align}
& \left< \chi (t) \middle| \left< \Phi (t) \middle|   
\hat{H}_{\textrm{elec}} \middle| \Phi (t) \right> \middle| \chi (t) \right> = \left< \Phi (t) \middle|   
\hat{H}_{\textrm{elec}} \middle| \Phi (t) \right>  \nonumber \\
& = E_0 +  \frac{1}{2} \sum_{ai,bj} 
\left(  \Theta_{ai}^\ast A_{ai,bj} \Theta_{bj}+ \Theta_{ai} A_{ai,bj} \Theta_{bj}^\ast \right. \nonumber \\
& \quad \quad \quad \quad  + \left. \Theta_{ai} B_{ai,bj} \Theta_{bj} + \Theta_{ai}^\ast B_{ai,bj} \Theta_{bj}^\ast  \right)
\end{align}
while the DSE contribution is 
\begin{align}
& \left< \chi (t) \middle| \left< \Phi (t) \middle|   
\frac{1}{2} \sum_{\alpha=1}^M \left( \bm{\lambda}_\alpha \cdot \left< \Delta \hat{\bm{\mu}} \right> \right)^2 \middle| \Phi (t) \right> \middle| \chi (t) \right> \nonumber \\
& =   \frac{1}{2}  \sum_{\alpha=1}^M  \left< \Phi (t) \middle|  
 \left( \bm{\lambda}_\alpha \cdot  \left<  \Delta  \hat{\bm{\mu}}  \right> \right)^2   \middle| \Phi (t) \right> \nonumber  \\
 & =   \frac{1}{2}  \sum_{\alpha, ai, bj}  \lambda_{ai}^\alpha  \lambda_{bj}^\alpha 
 \left< \Phi_0 \middle|   e^{\hat{\Lambda}_{\textrm{e}} }   \left( \hat{a}_i^\dagger \hat{a}_a +  \hat{a}_a^\dagger \hat{a}_i \right)  e^{-\hat{\Lambda}_{\textrm{e}} }    \middle| \Phi_0 \right>  \nonumber \\
 &    \times 
\left< \Phi_0 \middle|   e^{\hat{\Lambda}_{\textrm{e}} }   \left( \hat{a}_j^\dagger \hat{a}_b +  \hat{a}_b^\dagger \hat{a}_j \right)  e^{-\hat{\Lambda}_{\textrm{e}} }    \middle| \Phi_0 \right> \nonumber \\
& = \frac{1}{2}  \sum_{ai, bj}  \Delta_{ai,bj}
\left(  \Theta_{ai} (t)+  \Theta_{ai}^* (t)  \right)  \left(  \Theta_{bj} (t)+  \Theta_{bj}^* (t)  \right)
\end{align}
which uses the expressions in Eqs. \ref{eq:lambda-and-mu} and  \ref{eq:psi0-u-jb-ut-psi0}. 

The electron-photon coupling energy is 
\begin{align}
& - \left< \chi (t) \middle| \left< \Phi (t) \middle|   
\sum_{\alpha=1}^M \omega_\alpha \hat{q}_\alpha \left( \bm{\lambda}_\alpha \cdot \left< \Delta \hat{\bm{\mu}} \right> \right) \middle| \Phi (t) \right> \middle| \chi (t) \right> \nonumber \\
& = -  \sum_{\alpha=1}^M  \omega_\alpha \left< \chi(t) \middle|   \hat{q}_\alpha  \middle| \chi(t) \right>
\left< \Phi (t) \middle|  \left( \bm{\lambda}_\alpha \cdot  \Delta  \hat{\bm{\mu}} \right)  \middle| \Phi (t) \right>   \nonumber \\
& =  \sum_{\alpha, bj}   \sqrt{ \frac{\hbar \omega_\alpha}{2 }}  \left[ C_{\alpha} (t)   + C_{\alpha}^* (t) \right]  
\lambda_{bj}^\alpha  \left(  \Theta_{bj} (t)+  \Theta_{bj}^* (t)  \right) 
\end{align}
which uses the definition of photon coordinate in Eq. \ref{eq:q_alpha} and the expression in Eq. \ref{eq:b_and_b_dagger}. 

Finally, from Eq. \ref{eq:b_dagger_b}, the photon energy can be found to be  
\begin{align}
& \left< \chi (t) \middle| \left< \Phi (t) \middle|   
\sum_{\alpha=1}^M \left[ \frac{1}{2} \hat{p}_{\alpha}^2
+  \frac{1}{2} \omega_\alpha^2 \hat{q}_\alpha^2  \right]  \middle| \Phi (t) \right> \middle| \chi (t) \right> \nonumber \\
& = \sum_{\alpha=1}^M  \left< \chi(t) \middle|  
 \left[  \frac{1}{2} \hat{p}_{\alpha}^2 +  \frac{1}{2} \omega_\alpha^2 \hat{q}_\alpha^2 \right]   \middle| \chi(t) \right>  \nonumber \\
& =   \sum_{\alpha=1}^M  \hbar \omega_\alpha \left< \chi(t) \middle|  
 \left[  \hat{b}_\alpha^\dagger \hat{b}_\alpha + \frac{1}{2} \right]   \middle| \chi(t) \right>  \nonumber \\
& = \sum_{\alpha=1}^M  \hbar \omega_\alpha \left[ C_\alpha(t) C_\alpha^*(t) + \frac{1}{2} \right] 
\end{align}

Putting these together, we will get the following derivatives
\begin{eqnarray}
\frac {\partial E_{\textrm{PF}}} {\partial \Theta_{ai}}   
& = &  \sum_{bj} \left[ \left(A+\Delta\right)_{ai,bj}   \Theta_{bj}^\ast(t) + \left( B + \Delta \right) _{ai,bj}  \Theta_{bj}(t)    \right] \nonumber \\
&   & + \hbar g_{ai}^\alpha  \left[ C_{\alpha} (t)   + C_{\alpha}^* (t) \right]  \label{eq:orbital-rotation-derivative}  \\
\frac {\partial E_{\textrm{PF}}} {\partial C_\alpha}   
& = & \sum_{bj}   \hbar g_{bj}^\alpha  \left(  \Theta_{bj} (t)+  \Theta_{bj}^* (t)  \right) \nonumber \\
&   & +  \hbar \omega_\alpha C_\alpha^*(t)  
\label{eq:photon-phase-derivatives} 
\end{eqnarray}

\subsection{Equations of Motion} 
Let expand the Lagrangian to first-order 
\begin{eqnarray}
\mathcal{L} & = &   \left< \chi(t) \middle|  \left< \Phi (t) \middle| i \hbar \frac{\partial}{\partial t} - \hat{H}_{\textrm{PF}} \middle| \Phi (t) \right>  \middle| \chi(t) \right> \nonumber \\
& = &  i \sum_{ai} \Theta_{ai}^\ast (t) \frac{\partial}{\partial t} \Theta_{ai} (t)  + i \hbar \sum_{\alpha} C_{\alpha}^*(t) \frac{\partial}{\partial t} C_{\alpha}(t) \nonumber \\
&   & - E_{\textrm{PF}}  (\bm{\Theta}(t), \mathbf{C}(t)) 
\end{eqnarray}
The equation-of-motion for the orbital rotations is 
\begin{eqnarray}
- \frac{\partial}{\partial t} \left(  \frac{ \partial \mathcal{L} }{ \partial \dot{\Theta}_{ai} }  \right) & =  &  - \frac {\partial \mathcal{L}} {\partial \Theta_{ai}} 
\end{eqnarray}
namely,
\begin{eqnarray}
- i \hbar \frac{\partial }{\partial t}  \Theta_{ai}^\ast  & = &
\frac {\partial E_{\textrm{PF}}} {\partial \Theta_{ai}}  
\label{eq:eom-orbital-rotation} 
\end{eqnarray}

Let us parameterize the orbital rotations as
\begin{eqnarray}
\Theta_{ai} (t) = X_{ai} ~e^{ - i \Omega t} +   Y_{ai} ~e^{ i \Omega t} \\
\Theta_{ai}^\ast (t)=  X_{ai} ~e^{ i \Omega t}  +   Y_{ai} ~e^{ - i \Omega t}  
\end{eqnarray}
and photon phase change as 
\begin{eqnarray}
C_\alpha (t) = M_\alpha ~e^{ - i \Omega t} +   N_\alpha ~e^{ i \Omega t} \\
C_\alpha^\ast (t) = M_\alpha ~e^{ i \Omega t} +   N_\alpha ~e^{ -i \Omega t}
\end{eqnarray}

Plugging in orbital rotation derivatives in Eq. \ref{eq:orbital-rotation-derivative}, and separating the $e^{ i \Omega t}$ 
and $e^{ -i \Omega t}$ terms in Eq. \ref{eq:eom-orbital-rotation}, one gets the first two equations in Eq. \ref{eq:tddft-pf},
\begin{eqnarray}
& \Omega X_{ai} =  \sum_{bj} \left[ \left(A+\Delta\right)_{ai,bj} X_{bj} + \left(B+\Delta\right)_{ai,bj} Y_{bj} \right] \nonumber \\
& \quad \quad + \sum_\alpha \hbar g_{ai}^\alpha \left( M_{\alpha} + N_{\alpha} \right)  \\
&  - \Omega Y_{ai} =  \sum_{bj} \left[ \left(B+\Delta\right)_{ai,bj} X_{bj}+ \left(A+\Delta\right)_{ai,bj} Y_{bj}  \right] \nonumber \\
& \quad \quad + \sum_\alpha \hbar g_{ai}^\alpha   \left( M_{\alpha} + N_{\alpha} \right)  
\end{eqnarray}

The equation-of-motion for the photon phase is similar
\begin{eqnarray}
- i \hbar \frac{\partial }{\partial t}  C_\alpha^\ast  & = &
\frac {\partial E_{\textrm{PF}}} {\partial C_\alpha}  
\label{eq:eom-orbital-rotation-photon} 
\end{eqnarray}
which, upon the insertion of the derivatives in Eq. \ref{eq:photon-phase-derivatives}, leads to the last two equations in Eq. \ref{eq:tddft-pf},
\begin{eqnarray}
& \Omega M_\alpha  = \sum_{bj} \hbar g_{bj}^\alpha  \left( X + Y \right)_{bj}  + \hbar \omega_\alpha M_\alpha \\
& -\Omega N_\alpha  = \sum_{bj} \hbar g_{bj}^\alpha  \left( X + Y \right)_{bj}  + \hbar \omega_\alpha N_\alpha  
\end{eqnarray}

\section{2-State and 3-State JC Models}
\label{sec::three-state}
Let us consider a two-level system made of $\ket{e_1} \ket{0_\alpha}$ and  $\ket{g} \ket{1_\alpha}$, both of which have a resonance energy of $\hbar \omega$.  The 2-state JC equation is 
\begin{eqnarray} 
\begin{bmatrix}  \hbar \omega & s\hbar g_1 \\
s\hbar g_1 & \hbar \omega \end{bmatrix} 
\begin{bmatrix}  X_1 \\ M  \end{bmatrix} 
= \hbar \Omega \begin{bmatrix}   X_1 \\ M  \end{bmatrix} 
\label{eq:two-state-Hamiltonian} 
\end{eqnarray}    
where a scaling factor $s$ is added to track the perturbation order. 
Its solutions are known to be the lower and upper polaritons 
\begin{eqnarray} 
\ket{1_-}  & =  & \frac{1}{\sqrt{2}} \ket{e_1} \ket{0_\alpha} - \frac{1}{\sqrt{2}} \ket{g} \ket{1_\alpha}  \label{eq:lower-polariton-1}  \\
\ket{1_+}  & =  & \frac{1}{\sqrt{2}} \ket{e_1} \ket{0_\alpha} + \frac{1}{\sqrt{2}} \ket{g} \ket{1_\alpha}  \label{eq:upper-polariton-1} 
\end{eqnarray} 
with the energies being 
\begin{eqnarray}
\hbar \Omega_{1-} = \hbar \omega - s\hbar g_1  \label{eq:Omega1-C} \\
\hbar \Omega_{1+} = \hbar \omega + s\hbar g_1 \label{eq:Omega1+C}
\end{eqnarray} 
respectively.  

Now let us introduce a third state, $\ket{e_2} \ket{0_\alpha}$, which has an energy of $\hbar \omega_2$ and is well separated from the two polariton states.  The corresponding 3-state JC equation is   
\begin{eqnarray} 
\begin{bmatrix} \hbar \omega_2 & 0 & s\hbar g_2  \\
0 & \hbar\omega & s\hbar g_1 \\
s\hbar g_2 & s\hbar g_1 & \hbar \omega \end{bmatrix} 
\begin{bmatrix}  X_2 \\ X_1 \\ M  \end{bmatrix} 
= \hbar \Omega' \begin{bmatrix}  X_2 \\ X_1 \\ M  \end{bmatrix} 
\label{eq:TDA-JC-3-state} 
\end{eqnarray}
Upon switching to the basis of $\left[\ket{e_2} \ket{0_\alpha},~~~ \ket{1_+},  
~~~   \ket{1_-} \right]$, the 3-state JC equation becomes
\begin{align} 
\begin{bmatrix} \hbar \omega_2 &  \frac{s\hbar g_2}{\sqrt{2}} & -\frac{s\hbar g_2}{\sqrt{2}} \\
 \frac{s\hbar g_2}{\sqrt{2}} & \hbar \omega + s\hbar g_1  &  0  \\
-\frac{s\hbar g_2}{\sqrt{2}}  & 0  & \hbar \omega- s\hbar g_1 \end{bmatrix} 
\begin{bmatrix}  X_2 \\ P_{1+} \\ P_{1-}  \end{bmatrix} 
= \hbar \Omega' \begin{bmatrix}  X_2 \\ P_{1+} \\ P_{1-}  \end{bmatrix} 
\label{eq:TDA-JC-2} 
\end{align}

We can get the second-order perturbation to the energy of upper and lower polaritons as
\begin{align}
\hbar \Omega'_{1-} = \hbar \omega- s \hbar g_1 - \frac{   \frac{1}{2} s^2 \hbar^2 g_2^2} { \hbar \omega_2 - (\hbar \omega - s\hbar g_1)}  + \cdots \\
\hbar \Omega'_{1+} = \hbar \omega+ s\hbar g_1 - \frac{   \frac{1}{2} s^2 \hbar^2 g_2^2} { \hbar \omega_2 - (\hbar \omega + s\hbar g_1)} + \cdots
\end{align}
To the second order of $s$, we thus have 
\begin{align}
    \hbar \Omega'_{1-} = \hbar \omega- s\hbar g_1 -  s^2 \frac{ \hbar^2 g_2^2} { 2 (\hbar \omega_2 - \hbar \omega)}  + \mathcal{O} (s^3) \label{eq:Omega1-C2} \\
\hbar \Omega'_{1+} = \hbar \omega+ s\hbar g_1 - s^2  \frac{ \hbar^2 g_2^2} {2 ( \hbar \omega_2 - \hbar \omega)} + \mathcal{O} (s^3)  \label{eq:Omega1+C2}
\end{align}

Within the perturbation theory, the lower polariton wavefunction becomes 
\begin{align}
& \ket{1_-}'  =  \ket{1_-}^{(0)} + \ket{1_-}^{(1)} + \ket{1_-}^{(2)} + \cdots \nonumber \\
& \quad = \ket{ 1_- }  + \frac{- s \hbar g_2 / \sqrt{2} } {  ( \hbar \omega - s \hbar g_1) - \hbar \omega_2 } \ket{e_2} \ket{0_\alpha}  \nonumber \\
& \quad + \frac{ -\frac{1}{2} s^2 \hbar^2 g_2^2} {(-2 s\hbar g_1) \left[ (\hbar \omega - s\hbar g_1) - \hbar \omega_2 \right] }  \ket{1_+} + \cdots 
\end{align}
Plugging in the expressions in Eqs. \ref{eq:lower-polariton-1} and \ref{eq:upper-polariton-1} and truncating at the first order of $s$, one gets
\begin{eqnarray}
\ket{1_-}'  & = & -  \frac{1}{\sqrt{2}}  \left( 1 + \frac{s g_2^2}{ 4 g_1 ( \omega_2 -  \omega ) }  \right)  \ket{g} \ket{1_\alpha} \nonumber \\
&    & + \frac{1}{\sqrt{2}} \left( 1 -  \frac{s g_2^2}{ 4  g_1 ( \omega_2 - \omega ) }  \right)  \ket{e_1} \ket{0_\alpha} \nonumber \\
&    & + \frac{ s  g_2} {   \sqrt{2} (  \omega_2  -  \omega )  } \ket{e_2} \ket{0_\alpha}  + \mathcal{O} (s^2)   \label{eq:3-state-lower-wfn}
\end{eqnarray}
Similarly, the perturbed upper polariton wavefunction is 
\begin{eqnarray}
\ket{1_+}'  & = &   \frac{1}{\sqrt{2}}  \left( 1 - \frac{s g_2^2}{ 4 g_1 (\omega_2 -  \omega ) }  \right)  \ket{g} \ket{1_\alpha} \nonumber \\
&    & + \frac{1}{\sqrt{2}} \left( 1 + \frac{s  g_2^2}{ 4  g_1 ( \omega_2 -  \omega ) }  \right)  \ket{e_1} \ket{0_\alpha} \nonumber \\
&    & - \frac{ s  g_2} {   \sqrt{2} (  \omega_2  -  \omega )  } \ket{e_2} \ket{0_\alpha}  + \mathcal{O} (s^2)   \label{eq:3-state-upper-wfn}
\end{eqnarray}
The corresponding transition dipole moments are 
\begin{eqnarray}
\bm{\mu}_-' & = &  \bra{g} \bra{0_\alpha} \bm{\mu} \ket{1_+}'   = \frac{1}{\sqrt{2}} \left( 1 - \frac{s  g_2^2}{ 4  g_1 ( \omega_2 -  \omega ) }  \right) \bm{\mu}_1 \nonumber \\
& & +  \frac{ s  g_2} {   \sqrt{2} (  \omega_2  -  \omega )  } \bm{\mu}_2 + \mathcal{O} (s^2)  \label{eq:tm-lower} \\
\bm{\mu}_+' & = &  \bra{g} \bra{0_\alpha} \bm{\mu} \ket{1_+}'   = \frac{1}{\sqrt{2}} \left( 1 + \frac{s  g_2^2}{ 4  g_1 ( \omega_2 -  \omega ) }  \right) \bm{\mu}_1 \nonumber \\
& & -  \frac{ s  g_2} {   \sqrt{2} (  \omega_2  -  \omega )  } \bm{\mu}_2 + \mathcal{O} (s^2)  \label{eq:tm-upper} 
\end{eqnarray}

If larger coupling occurs with the second-excited state of the gas-phase molecule  ($g_2 \gg g_1$), the upper polariton $\ket{1_+}$ (at small $s$ values) gradually loses photon contributions as the $s$ values increases and the coupling between other states becomes more dominant.  In the limit that $M$ approaches zero, it corresponding eigenvector in Eq. \ref{eq:TDA-JC-3-state} becomes $(X_2, X_1, 0)^\mathrm{T}$. The third equation of Eq. \ref{eq:TDA-JC-3-state} requires that $X_2 g_2 + X_1 g_1 = 0$ or $X_2 / X_1 = - g_1 / g_2$. 
For this state without any photon character, 
the normalized eigenvector is thus
\begin{align}
\lim_{g_2 \gg g_1,s \gg 0}  \ket{1_+}'   
= \frac{g_2}{ g_1^2 + g_2^2 }  \ket{e_1} \ket{0_\alpha}  
- \frac{g_1}{ g_1^2 + g_2^2 } \ket{e_2} \ket{0_\alpha}  
\end{align}
with the energy being 
\begin{eqnarray}
\lim_{g_2 \gg g_1,s \gg 0} \Omega'_{1_+} = \omega + 
\frac{g_1^2 (\omega_2 - \omega)} { g_1^2 + g_2^2}.    \label{eq:Omega1+-3-state-limit}
\end{eqnarray} 

\bibliography{bibliography}
\end{document}


\preprint{AIP/123-QED}

\title{Supporting Information for: Quantum-Electrodynamical Time-Dependent Density Functional Theory. I. A Gaussian Atomic Basis Implementation}

\author{Junjie Yang}
\affiliation{Department of Chemistry and Biochemistry, University of Oklahoma, Norman, Oklahoma 73019, USA}

\author{Qi Ou}
\email{qiou@tsinghua.edu.cn}
\affiliation{MOE Key Laboratory of Organic OptoElectronics and Molecular Engineering, Department of Chemistry, Tsinghua
University, Beijing 100084, China.}

\author{Zheng Pei}
\affiliation{State Key Laboratory of Physical Chemistry of Solid Surfaces, Collaborative Innovation Center of Chemistry for Energy Materials, 
Fujian Provincial Key Laboratory of Theoretical and Computational Chemistry, and Department of Chemistry, College of Chemistry and Chemical Engineering, Xiamen University, Xiamen 361005, P. R. China.}

\author{Hua Wang} 
\affiliation{Homer L. Dodge Department of Physics and Astronomy, University of Oklahoma, Norman, Oklahoma 73019, USA} 

\author{Binbin Weng} 
\affiliation{Microfabrication Research and Education Center and School of Electrical and Computer Engineering, University of Oklahoma, Norman, Oklahoma 73019, USA}

\author{Zhigang Shuai}
\email{zgshuai@tsinghua.edu.cn}
\affiliation{MOE Key Laboratory of Organic OptoElectronics and Molecular Engineering, Department of Chemistry, Tsinghua
University, Beijing 100084, China.}

\author{Kieran Mullen}
\email{kieran@ou.edu}
\affiliation{Homer L. Dodge Department of Physics and Astronomy, University of Oklahoma, Norman, Oklahoma 73019, USA} 

\author{Yihan Shao}
\email{yihan.shao@ou.edu}
\affiliation{Department of Chemistry and Biochemistry, University of Oklahoma, Norman, Oklahoma 73019, USA}

\date{\today}
             
\maketitle

\section{Computation of Absorption Spectra}
The first quantity used to calculate the photoabsorption is the oscillation strength,\cite{flick_lightmatter_2019} (within atomic unit, \emph{i.e.}, $\hbar = 1$)
\begin{equation}
    f_I = \frac{2}{3} \omega_I |\boldsymbol{\mu}_I|^2
\end{equation}
where the transition dipole moment is defined as,
\begin{equation}
    \boldsymbol{\mu}_I = \bra{\Phi_0} \hat{\boldsymbol{\mu}} \ket{\Psi_I}
\end{equation}
which is calculated as,
\begin{eqnarray}
\boldsymbol{\mu}_I = \sum_{ai} X^I_{ai} \bm{\mu}_{ai}
\end{eqnarray} 
for QED-TDA models and 
\begin{eqnarray}
\boldsymbol{\mu}_I = \sum_{ai} \left( X^I_{ai} + Y^I_{ai} \right) \bm{\mu}_{ai}
\end{eqnarray} 
for QED-TDDFT models, both using the mo-based dipole in Eq. 1 of the main text.  
We apply a Lorentzian broadening of the following form,
\begin{equation}
\Gamma\left(\omega, \omega_{I}\right)=\frac{1}{\pi} \frac{\Delta}{\left(\omega-\omega_{I}\right)^{2}+\Delta^{2}}
\end{equation}
where $\omega_{I}$ is the excitation energy, and $\Delta$ is the broadening parameter. The actual dipole strength function is then obtained by,
\begin{equation}
S(\omega)=\sum_{I} f_{I} \Gamma\left(\omega, \omega_{I}\right)
\end{equation}
We obtain the spectra for systems not immersed in the photon bath in this paper, where the peaks have been broadened by applying the broadening parameter as with $\Delta = 10^{-2} \, \mathrm{eV}$.

\newpage
\section{Optimized Ground-State Geometries of Ethene, Formaldehyde, and Benzaldehyde using PBE0 Functional and 6-311++G** Basis Set}

\subsection{Optimized Ground-State Geometries of Ethene Molecule}
\begin{verbatim}
C2H4 PBE0 6-311++G** 
6
-78.5042464024453
C  0.0000000000  0.6662105645  0.0000000000
C  0.0000000000 -0.6662105645  0.0000000000
H  0.0000000000  1.2386216337  0.9292861049
H  0.0000000000  1.2386216337 -0.9292861049
H  0.0000000000 -1.2386216337  0.9292861049
H  0.0000000000 -1.2386216337 -0.9292861049
\end{verbatim}

\subsection{Optimized Ground-State Geometries of Formaldehyde Molecule}
\begin{verbatim}
CH2O PBE0 6-311++G** 
4
-114.406100317565
O  0.0000000000   0.0000000000  0.6881243879
C  0.0000000000   0.0000000000 -0.5150066070
H  0.0000000000   0.9462463671 -1.1029423155
H  0.0000000000  -0.9462463671 -1.1029423155  
\end{verbatim}

\subsection{Optimized Ground-State Geometries of Benzaldehyde Molecule}
\begin{verbatim}
C6H5CHO PBE0 6-311++G** 
14
-345.255583702600
C   0.0444176449 -1.1018291187 0.0000000000 
C  -1.3261410267 -1.3281816235 0.0000000000 
C  -2.2133752765 -0.2471903340 0.0000000000 
C  -1.7312839975  1.0611346821 0.0000000000 
C  -0.3570221851  1.2888292340 0.0000000000 
C   0.5334742585  0.2107166850 0.0000000000 
C   1.9887836264  0.4668394275 0.0000000000 
O   2.8410453615 -0.3945408322 0.0000000000 
H  -1.7117538780 -2.3482451649 0.0000000000 
H  -3.2892064941 -0.4282442776 0.0000000000 
H  -2.4267559760  1.9009365066 0.0000000000 
H   0.0335635546  2.3089090226 0.0000000000 
H   2.2739453181  1.5465885188 0.0000000000 
H   0.7587263192 -1.9255316626 0.0000000000 
\end{verbatim}

\newpage
\section{Numerical Results}
\subsection{Ethene}
\begin{table}[h!]
\tabcolsep 12pt
\caption{The ethene (PBE0/6-311++G**) gas-phase TDA and TDDFT excited state energies and transition dipoles.}
\begin{tabular}{lrrrrrrrrr}
\toprule
 & \multicolumn{4}{c}{TDA} & & \multicolumn{4}{c}{TDDFT} \\ \cline{2-5}   \cline{7-10} 
State & $E$(eV) & $\mu_x$ (au) & $\mu_y$ (au) & $\mu_z$ (au) & & $E$ (eV) & $\mu_x$(au) & $\mu_y$(au) & $\mu_z$(au) \\ \midrule
1   &  6.961  &  0.589       & -0.000       & -0.000     &  &  6.960       &  0.585      & -0.000      & -0.000  \\
2   &  7.563  & -0.000       &  0.000       & -0.000     &  &  7.562       & -0.000      & -1.357      &  0.000  \\
3   &  7.595  & -0.000       &  0.000       &  0.000     &  &  7.562       & -0.000      &  0.000      &  0.000  \\
4   &  8.003  & -0.000       & -1.531       & -0.000     &  &  7.591       &  0.000      & -0.000      & -0.000  \\
5   &  8.062  &  0.000       &  0.000       &  0.000     &  &  8.044       &  0.000      &  0.000      &  0.000  \\
6   &  8.671  &  0.000       & -0.000       &  0.000     &  &  8.671       &  0.000      & -0.000      & -0.000  \\
7   &  8.843  &  0.000       & -0.000       &  0.000     &  &  8.804       & -0.000      & -0.000      & -0.000  \\
8   &  9.073  &  0.068       &  0.000       & -0.000     &  &  9.072       & -0.065      & -0.000      &  0.000  \\
9   &  9.082  &  0.000       &  0.000       &  0.000     &  &  9.078       &  0.000      &  0.000      &  0.000  \\
10   &  9.626  & -0.000       &  0.000       &  0.000    &  &  9.604       & -0.000      & -0.000      & -0.000  \\
11   &  9.701  & -0.000       & -0.000       &  0.640    &  &  9.696       & -0.000      &  0.000      & -0.628  \\
12   &  9.771  &  0.000       & -0.587       & -0.000    &  &  9.760       &  0.000      & -0.588      &  0.000  \\
13   & 10.022  & -1.197       & -0.000       & -0.000    &  &  9.992       & -1.156      &  0.000      &  0.000  \\
14   & 10.535  &  0.000       &  0.000       & -0.000    &  & 10.534       & -0.000      & -0.000      & -0.000  \\
15   & 10.741  &  0.000       & -0.000       & -0.000    &  & 10.625       & -0.000      &  0.193      & -0.000  \\
16   & 10.795  &  0.000       &  0.000       &  0.000    &  & 10.739       & -0.000      & -0.000      & -0.000  \\
17   & 10.848  &  0.000       & -0.000       & -0.000    &  & 10.794       &  0.000      & -0.000      &  0.000  \\
18   & 10.857  &  0.000       &  0.435       & -0.000    &  & 10.840       & -0.000      &  0.000      & -0.000  \\
19   & 10.967  &  0.002       &  0.000       & -0.000    &  & 10.961       &  0.018      &  0.000      &  0.000  \\
20   & 10.997  & -0.000       &  0.000       & -0.000    &  & 10.992       &  0.000      & -0.000      & -0.000  \\
 
\bottomrule
\end{tabular}
\end{table}
\newpage

\subsection{Formaldehyde}
\begin{table}[h!]
\tabcolsep 13pt
\caption{The formaldehyde (PBE0/6-311++G**) gas-phase TDA and TDDFT excited state energies and transition dipoles.}
\begin{tabular}{lrrrrrrrrr}
\toprule
 & \multicolumn{4}{c}{TDA} & & \multicolumn{4}{c}{TDDFT} \\ \cline{2-5}   \cline{7-10} 
State & $E$(eV) & $\mu_x$ (au) & $\mu_y$ (au) & $\mu_z$ (au) & & $E$ (eV) & $\mu_x$(au) & $\mu_y$(au) & $\mu_z$(au) \\ \midrule
1   &  3.994  &  0.000       &  0.000       & -0.000       & &  3.966       &  0.000      &  0.000      & -0.000  \\
2   &  6.784  &  0.490       &  0.000       &  0.000       & &  6.777       &  0.475      &  0.000      &  0.000  \\
3   &  7.760  & -0.000       &  0.000       & -0.544       & &  7.748       &  0.000      &  0.000      &  0.527  \\
4   &  7.857  & -0.373       &  0.000       &  0.000       & &  7.851       & -0.367      &  0.000      &  0.000  \\
5   &  8.563  & -0.000       &  0.000       & -0.000       & &  8.562       &  0.000      & -0.000      &  0.000  \\
6   &  9.243  &  0.000       & -0.063       &  0.000       & &  9.165       & -0.000      &  0.051      &  0.000  \\
7   &  9.672  & -0.000       & -0.000       & -0.545       & &  9.480       & -0.000      & -0.000      & -0.685  \\
8   &  9.898  &  0.254       & -0.000       & -0.000       & &  9.890       & -0.252      &  0.000      &  0.000  \\
9   & 10.209  &  0.000       &  0.000       &  0.000       & & 10.130       & -0.000      &  0.000      &  0.372  \\
10   & 10.400  &  0.000       & -0.000       & -0.541      & & 10.196       & -0.000      & -0.000      & -0.000  \\
11   & 10.438  &  0.000       &  0.299       & -0.000      & & 10.437       & -0.000      & -0.299      & -0.000  \\
12   & 10.645  &  0.674       & -0.000       &  0.000      & & 10.606       &  0.627      &  0.000      &  0.000  \\
13   & 11.380  & -0.000       & -0.000       & -0.000      & & 11.379       &  0.000      & -0.000      &  0.000  \\
14   & 11.556  &  0.000       &  0.409       & -0.000      & & 11.549       &  0.000      &  0.401      &  0.000  \\
15   & 11.709  &  0.017       &  0.000       &  0.000      & & 11.706       & -0.024      &  0.000      & -0.000  \\
16   & 11.933  &  0.000       & -0.000       & -0.056      & & 11.924       &  0.000      & -0.000      & -0.073  \\
17   & 12.407  &  0.000       & -0.000       &  0.182      & & 12.357       &  0.000      & -0.000      & -0.285  \\
18   & 12.563  &  0.000       &  0.000       & -0.000      & & 12.560       & -0.000      & -0.000      &  0.000  \\
19   & 12.851  & -0.000       &  0.000       &  0.912      & & 12.609       & -0.000      &  0.000      &  0.988  \\
20   & 12.852  & -0.472       &  0.000       & -0.000      & & 12.848       & -0.477      & -0.000      & -0.000  \\
\bottomrule
\end{tabular}
\end{table}

\begin{table}[h!] 
\tabcolsep 13pt
\caption{The formaldehyde upper($E_{+}$)/lower($E_{-}$) polariton energy (in eV) and the Rabi splitting ($\Delta E$) using different QED-TDDFT models with respect to the coupling strength $\lambda$.}
\begin{tabular}{clccccccccc}
\toprule
$\lambda$ & Energy & \multicolumn{4}{c}{TDA} & & \multicolumn{4}{c}{TDDFT}                           \\ \cline{3-6}  \cline{8-11}
(au) &   & JC     & Rabi   & RWA    & PF & & \multicolumn{1}{c}{JC} & Rabi   & RWA    & PF     \\ \hline
\multirow{3}{*}{ 0.045} &  $E_+$ &  6.942 &  6.942 &  6.945 &  6.944 & &  6.921                 &  6.921 &  6.923 &  6.923 \\
&  $E_-$ &  6.511 &  6.508 &  6.524 &  6.521 & &  6.494 &  6.491 &  6.509 &  6.505 \\
&  $\Delta E$ &  0.431 &  0.434 &  0.421 &  0.423 & &  0.427                 &  0.430 &  0.414 &  0.417 \\\hline
\multirow{3}{*}{ 0.067} &  $E_+$ &  6.989 &  6.989 &  6.993 &  6.992 & & 6.960 & 6.959 &  6.962 &  6.961 \\
&  $E_-$ &  6.328 &  6.318 &  6.366 &  6.356 & &  6.290                 &  6.278 &  6.335 &  6.324 \\
&  $\Delta E$ &  0.661 &  0.671 &  0.627 &  0.636 & & 0.669                 &  0.681 &  0.627 &  0.637 \\\hline
\multirow{3}{*}{ 0.090} &  $E_+$ &  7.022 &  7.021 &  7.026 &  7.025 & &  6.985                 &  6.984 &  6.987 &  6.987 \\
&  $E_-$ &  6.116 &  6.092 &  6.196 &  6.177 & &  6.044                 &  6.013 &  6.146 &  6.122 \\
&  $\Delta E$ &  0.906 &  0.929 &  0.830 &  0.848 & & 0.940                 &  0.972 &  0.841 &  0.865 \\
\bottomrule
\end{tabular}
\label{tab::ch2o}
\end{table}
\newpage

\subsection{Benzaldehyde}
\hspace{-1.0\textwidth}
\begin{table}[h!]
\tabcolsep 13pt
\caption{The benzaldehyde (PBE0/6-311++G**) gas-phase TDA and TDDFT excited state energies and transition dipoles.}
\begin{tabular}{lrrrrrrrrr}
\toprule
 & \multicolumn{4}{c}{TDA} & & \multicolumn{4}{c}{TDDFT} \\ \cline{2-5}   \cline{7-10} 
State & $E$(eV) & $\mu_x$ (au) & $\mu_y$ (au) & $\mu_z$ (au) & & $E$ (eV) & $\mu_x$(au) & $\mu_y$(au) & $\mu_z$(au) \\ \midrule
1   &  3.722  &  0.000       &  0.000       & -0.033       & &  3.699       &  0.000      & -0.000      & -0.037  \\
2   &  4.879  & -0.104       &  0.414       &  0.000       & &  4.810       & -0.065      &  0.422      & -0.000  \\
3   &  5.488  & -1.492       & -0.271       & -0.000       & &  5.253       & -1.390      & -0.213      &  0.000  \\
4   &  5.759  & -0.000       & -0.000       & -0.024       & &  5.758       & -0.000      &  0.000      & -0.021  \\
5   &  6.548  &  0.628       & -0.689       &  0.000       & &  6.342       & -0.650      &  0.809      & -0.000  \\
6   &  6.683  &  0.000       &  0.000       &  0.036       & &  6.620       & -1.283      & -0.534      &  0.000  \\
7   &  6.700  &  0.000       &  0.000       &  0.165       & &  6.675       &  0.000      & -0.000      &  0.036  \\
8   &  6.714  & -0.000       &  0.000       &  0.100       & &  6.699       &  0.000      & -0.000      &  0.174  \\
9   &  6.741  &  0.039       &  0.581       & -0.000       & &  6.704       &  0.000      & -0.000      &  0.064  \\
10   &  6.781  &  0.000       & -0.000       & -0.062      & &  6.756       &  0.911      & -0.149      & -0.000  \\
11   &  6.971  & -1.759       & -0.279       & -0.000      & &  6.780       &  0.000      & -0.000      &  0.064  \\
12   &  7.075  & -0.354       & -0.478       & -0.000      & &  7.069       &  0.210      &  0.422      & -0.000  \\
13   &  7.153  &  0.000       &  0.000       &  0.076      & &  7.141       &  0.000      & -0.000      &  0.065  \\
14   &  7.188  & -0.000       &  0.000       &  0.196      & &  7.187       & -0.000      &  0.000      &  0.199  \\
15   &  7.218  & -0.000       &  0.000       &  0.332      & &  7.217       & -0.000      &  0.000      & -0.323  \\
16   &  7.311  &  0.034       &  0.140       & -0.000      & &  7.309       & -0.020      & -0.114      & -0.000  \\
17   &  7.431  &  0.000       & -0.000       &  0.232      & &  7.429       &  0.000      & -0.000      & -0.224  \\
18   &  7.539  &  0.000       & -0.000       &  0.029      & &  7.521       & -0.185      & -0.715      & -0.000  \\
19   &  7.621  & -0.589       & -0.478       & -0.000      & &  7.539       &  0.000      & -0.000      &  0.031  \\
20   &  7.714  &  0.613       & -0.813       &  0.000      & &  7.581       &  0.627      & -0.935      & -0.000  \\
\bottomrule
\end{tabular}
\end{table}

\begin{figure}[htp]
    \centering
    \includegraphics[width=0.8\linewidth]{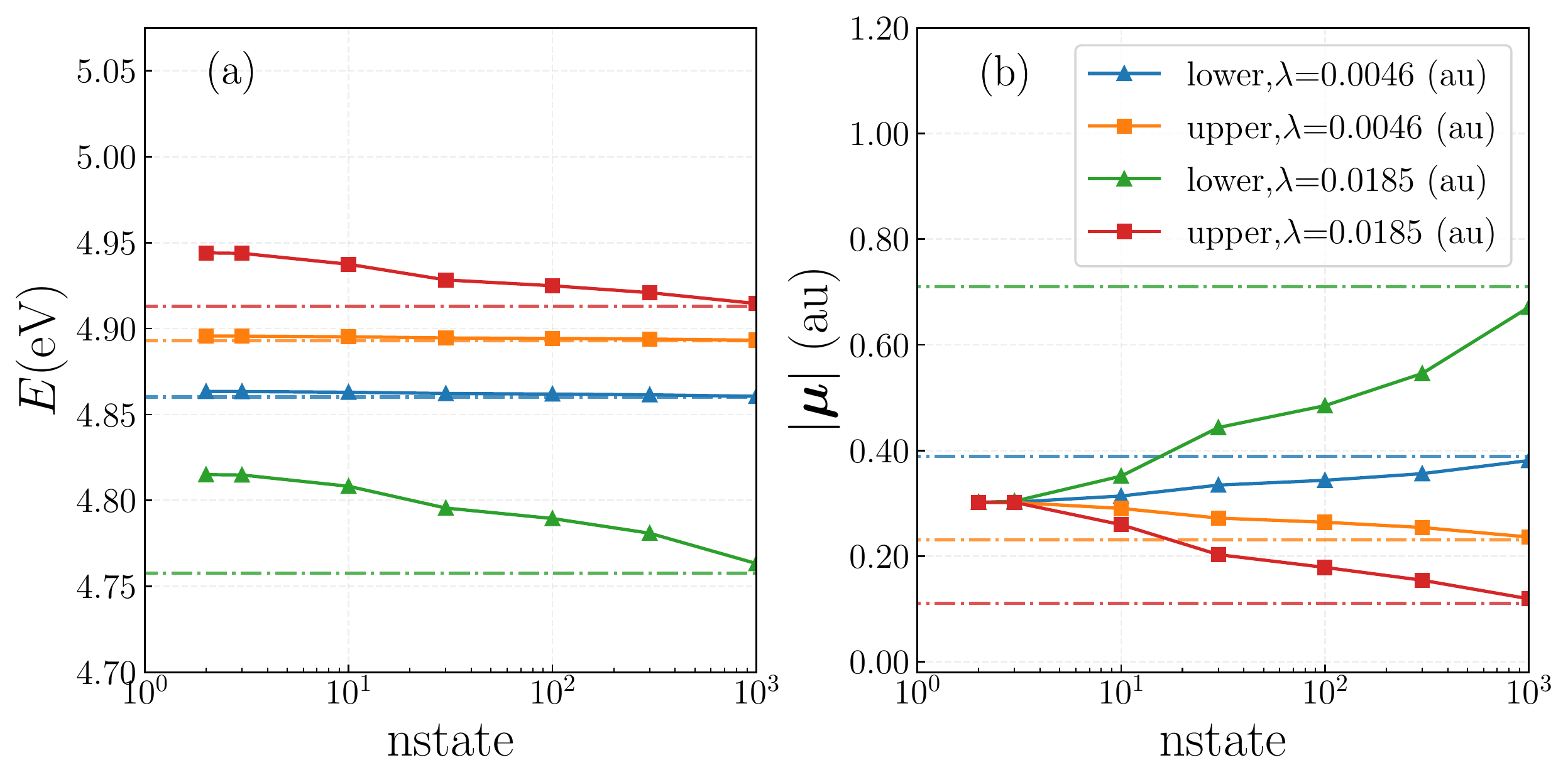}
    \caption{(a) Energy of the benzaldehyde lower and upper polaritons at different coupling strengths and (b) the corresponding transition dipole moments with respect to the number of included gas-phase TDA states in the Jay-Cummings model Hamiltonian.  Dashed lines indicate the TDA-JC values.}
    \label{fig::c6h5cho-2}
\end{figure}

\begin{table}[h!]
\tabcolsep 6pt
\caption{The benzaldehyde resonant upper($E_{+}$) / lower($E_{-}$) polariton energy (in eV)  and off-resonant upper($E^\prime_{+}$) / lower($E^\prime_{-}$) polariton energy (in eV) using different QED-TDDFT models with respect to the coupling strength $\lambda$.}
\begin{tabular}{clcccccllcccc}
\hline
$\lambda$            & Energy & \multicolumn{5}{c}{TDA}                     &  & \multicolumn{5}{l}{TDDFT}                   \\ \cline{3-7} \cline{9-13} 
(au)                 & (eV)   & 2-state & JC     & Rabi   & RWA    & PF     &  & 2-state & JC     & Rabi   & RWA    & PF     \\ \hline
          & $E_-$  & 4.8794  & 4.8794 & 4.8794 & 4.8794 & 4.8794 &  & 4.8103  & 4.8103 & 4.8103 & 4.8103 & 4.8103 \\
0.0000 $\,$           & $E_+$  & 4.8794  & 4.8794 & 4.8794 & 4.8794 & 4.8794 &  & 4.8103  & 4.8103 & 4.8103 & 4.8103 & 4.8103 \\ \cline{2-13} 
\multicolumn{1}{l}{} & $E^\prime_-$  & 4.8794  & 4.8794 & 4.8794 & 4.8794 & 4.8794 &  & 4.8103  & 4.8103 & 4.8103 & 4.8103 & 4.8103 \\
\multicolumn{1}{l}{} & $E^\prime_+$  & 4.8994  & 4.8994 & 4.8994 & 4.8994 & 4.8994 &  & 4.8303  & 4.8303 & 4.8303 & 4.8303 & 4.8303 \\ \hline
          & $E_-$  & 4.8633  & 4.8602 & 4.8602 & 4.8603 & 4.8603 &  & 4.7943  & 4.7909 & 4.7909 & 4.7910 & 4.7910 \\
0.0046 $\,$           & $E_+$  & 4.8955  & 4.8930 & 4.8930 & 4.8930 & 4.8930 &  & 4.8263  & 4.8235 & 4.8235 & 4.8236 & 4.8235 \\ \cline{2-13} 
\multicolumn{1}{l}{} & $E^\prime_-$  & 4.8704  & 4.8689 & 4.8689 & 4.8690 & 4.8690 &  & 4.8014  & 4.7997 & 4.7997 & 4.7998 & 4.7998 \\
\multicolumn{1}{l}{} & $E^\prime_+$  & 4.9084  & 4.9042 & 4.9042 & 4.9042 & 4.9042 &  & 4.8392  & 4.8347 & 4.8347 & 4.8347 & 4.8347 \\ \hline
          & $E_-$  & 4.8471  & 4.8341 & 4.8340 & 4.8347 & 4.8345 &  & 4.7782  & 4.7639 & 4.7638 & 4.7645 & 4.7644 \\
0.0093 $\,$           & $E_+$  & 4.9117  & 4.9023 & 4.9023 & 4.9023 & 4.9023 &  & 4.8423  & 4.8323 & 4.8323 & 4.8324 & 4.8324 \\ \cline{2-13} 
\multicolumn{1}{l}{} & $E^\prime_-$  & 4.8556  & 4.8458 & 4.8457 & 4.8464 & 4.8463 &  & 4.7867  & 4.7759 & 4.7758 & 4.7766 & 4.7764 \\
\multicolumn{1}{l}{} & $E^\prime_+$  & 4.9232  & 4.9104 & 4.9104 & 4.9104 & 4.9104 &  & 4.8538  & 4.8401 & 4.8401 & 4.8402 & 4.8402 \\ \hline
          & $E_-$  & 4.8310  & 4.8001 & 4.7996 & 4.8019 & 4.8015 &  & 4.7622  & 4.7281 & 4.7276 & 4.7301 & 4.7296 \\
0.0139 $\,$           & $E_+$  & 4.9278  & 4.9086 & 4.9086 & 4.9087 & 4.9087 &  & 4.8583  & 4.8381 & 4.8381 & 4.8382 & 4.8381 \\ \cline{2-13} 
\multicolumn{1}{l}{} & $E^\prime_-$  & 4.8400  & 4.8135 & 4.8130 & 4.8154 & 4.8150 &  & 4.7712  & 4.7419 & 4.7414 & 4.7440 & 4.7435 \\
\multicolumn{1}{l}{} & $E^\prime_+$  & 4.9388  & 4.9147 & 4.9147 & 4.9147 & 4.9147 &  & 4.8694  & 4.8438 & 4.8438 & 4.8439 & 4.8439 \\ \hline
          & $E_-$  & 4.8149  & 4.7576 & 4.7564 & 4.7621 & 4.7610 &  & 4.7462  & 4.6827 & 4.6813 & 4.6878 & 4.6865 \\
0.0185 $\,$           & $E_+$  & 4.9440  & 4.9129 & 4.9129 & 4.9130 & 4.9130 &  & 4.8743  & 4.8419 & 4.8419 & 4.8420 & 4.8420 \\ \cline{2-13} 
\multicolumn{1}{l}{} & $E^\prime_-$  & 4.8241  & 4.7721 & 4.7709 & 4.7768 & 4.7756 &  & 4.7554  & 4.6976 & 4.6962 & 4.7028 & 4.7015 \\
\multicolumn{1}{l}{} & $E^\prime_+$  & 4.9547  & 4.9176 & 4.9176 & 4.9176 & 4.9176 &  & 4.8851  & 4.8462 & 4.8462 & 4.8463 & 4.8462 \\ \hline
\end{tabular}
\end{table}
\newpage

\bibliography{bibliography}